\documentclass[aps,preprint,showpacs,preprintnumbers,amsmath,amssymb]{revtex4}
\usepackage[dvips]{graphicx}
\usepackage{float}
\begin{document}

\title{Total probability and number of particles for fermion production in external electric and magnetic fields in  de Sitter universe}
\author{Mihaela-Andreea B\u aloi \thanks{E-mail:~mihaela.baloi88@gmail.com}\\
{\small \it Politehnica University of Timi\c soara,}\\
{\small \it V. Parvan Ave. 2 RO-300223 Timi\c soara,  Romania}
\\Diana Popescu and Cosmin Crucean \thanks{E-mail:~~crucean@physics.uvt.ro}\\
{\small \it Faculty of Physics, West University of Timi\c soara,}\\
{\small \it V. Parvan Ave. 4 RO-300223 Timi\c soara,  Romania}}

\begin{abstract}
In this paper we present a method for computing the total probability corresponding to the processes of fermion pair production in dipole magnetic field and external Coulomb field in a de Sitter geometry. For that we rewrite the functions that define the amplitudes in terms of oscillatory functions and we use the Taylor expansion of these functions for arriving at the final form of the momenta integrals.
The total probability is analysed in terms of the ratio between the mass of the fermion and expansion parameter and we recover the Minkowski limit when the expansion parameter is vanishing. Finally it is proven that from our results we can compute the number of fermions in terms of expansion parameter.
\end{abstract}

\maketitle
\section{Introduction}
The first studies that attest the importance of the perturbative treatment of the problem of particle production were done some time ago \cite{17,27,28,29}.
In the last years a perturbative method for studying the phenomenon of particle production in Robertson-Walker spacetimes was developed \cite{6,7,8,9,10,11,12,13,14,23} and this is based on the using of the exact solutions of the free field equations \cite{7,9,cc}, for computations related to transition amplitudes and probabilities. This method is based on the S matrix approach and helps one to compute the first order perturbative processes corresponding to the de Sitter QED. In other words in this metric the processes that generate particle production have no longer vanishing amplitudes due to the fact that the translational invariance with respect to time is lost \cite{9,12}. These studies were done in the Coulomb gauge which is the only gauge suitable for study the free electromagnetic field since the Lorentz condition becomes conformal invariant only in this gauge in de Sitter geometry \cite{7,10}. Then using the conformal invariance of the Maxwell equations, the plane wave solutions were written in the chart $\{t_{c},\vec{x}\,\}$ \cite{9}, and in addition one can establish the form of the external electromagnetic fields that are directly related to the problem of particle production \cite{10}. The basic results obtained until now establish that the momentum conservation law is broken when one studies interactions with an external electromagnetic field in de Sitter geometry \cite{10,12,13} and that the phenomenon of pair generation is possible only in early universe. The transitions which refer to the emission and absorbtion processes together with the vacuum transitions preserve the momentum conservation law \cite{7,8}, and for these processes it was proven that the transition rate could be established in a de Sitter geometry \cite{rt}. Contrary to this when the momentum is no longer conserved we do not have arguments for defining the transition rate. In these circumstances the total probability is the quantity that gives complete information about the processes of pair production in external electromagnetic fields. But to obtain this quantity from the transition amplitude requires, in most cases mathematical tools that were not used elsewhere or integrals that are not known in literature. All the transition amplitudes with fermions in external field have a complicated dependence on momenta, given in terms of polynomial factors and hypergeometric functions and this further complicates the study of the total probability.

The problem of particle production in magnetic fields in a curved space-time was studied only recently and the main results can be found in \cite{10, 12, bs}. In \cite{10} the problem of fermion production in dipole magnetic field on de Sitter geometry was studied by computing the nonvanishing first order transition amplitude of de Sitter QED. The nonperturbative approach to the problem of pair production in a uniform magnetic field on de Sitter geometry was done in \cite{bs} and has as result the number of created pairs per unit of volume. In this paper we want to extend our results obtained in \cite{10,13} for fermion production in external electromagnetic fields by computing the momenta integrals that give the total probabilities.

In our study we use the transition amplitudes and probability densities derived in \cite{10} and \cite{13}, for fermion pair production in dipole magnetic field and Coulomb field in de Sitter spacetime. The problem of scalar pair production in dipole magnetic field was studied in \cite{12}, and the total probability was computed, proving that this quantity is important only for large expansion conditions \cite{12}. However in the case of the Dirac field the analytical structure of the amplitude is rather complicated and in \cite {10} and \cite{13} the expressions for the total probabilities were not computed. In the present paper we want to extend the results obtained in \cite{12} to the case of fermion production in dipole magnetic field and Coulomb field.  In the case of the Dirac field we specify that the problem of origin of magnetic fields in universe was studied in \cite{5}. The problem of particle generation in strong magnetic fields in Minkowski space-time was studied by using nonperturbative methods \cite{3,4} and have as result the density number of produced particles.

The starting point of our analysis will be the analytical expression for the transition amplitudes which can be written in terms of trigonometric functions without making any approximations. Then these expressions can be used for explaining the oscillatory behaviour of the  probability density obtained in \cite{10,13}. In the second section we present the main steps for obtaining the analytical expression of transition amplitude and probability density in terms of trigonometric functions in the case of fermion production in dipole magnetic field and in the end we compute the total probability. In section three we obtain the analytical formulas for the total probability of fermion production in Coulomb field on de Sitter geometry. In section four we obtain the number of particles and we discuss the dependence of this quantity in terms of expansion parameter and fermion mass. Section five is dedicated discussions related to our results. We use natural units such that $\hbar=1\,,c=1$.

\section{Pair production in dipole magnetic field}
We start with the result obtained in \cite{10}, for the transition amplitude of fermion pair generation in dipole magnetic field \cite{1}, on de Sitter spacetime \cite{0}. Only the case $p>p'$ will be taken into account and we specify that for $p'>p$, the analysis is similar. The final equation for the transition amplitude in the first order of perturbation theory that describe fermion pair generation in dipole magnetic field in de Sitter space-time is \cite{10}:
\begin{equation}\label{an1}
\mathcal{A}_{e^{-}e^{+}}= -\frac{ie}{4\pi^{3}|\vec{p}+\vec{p}\,'|^{2}}\frac{\theta(p-p')}{p}\left[f_{k}^{*}\left(\frac{p'}{p}\right)-sgn(\sigma\sigma')f_{k}\left(\frac{p'}{p}\right)\right]
\xi_{\sigma}^{+}(\vec{p}\,)[\vec{\sigma}\cdot(\vec{\mathcal{M}}\times(\vec{p}+\vec{p}\,'))]\eta_{\sigma'}(\vec{p}\,'),
\end{equation}
where $\theta(p-p')$ is the unit step function, $\vec{\mathcal{M}}$ is the magnetic dipole moment and the functions $f_k\left(\frac{p'}{p}\right)$ are given by \cite{10}:
\begin{equation}\label{f}
f_{k}^{*}\left(\frac{p'}{p}\right)= \frac{\pi}{\cosh{(\pi k)}}\left[\left(\frac{p'}{p}\right)^{-ik}-\left(\frac{p'}{p}\right)^{1+ik}\right]\left(1-\left(\frac{p'}{p}\right)^{2}\right)^{-1}.
\end{equation}
The functions $f_k\left(\frac{p'}{p}\right)$ depend on the parameter (fermion mass/expansion factor) denoted by $k=m/\omega$, which is the key factor that contains the space expansion contribution to our amplitude. The study in \cite{10} was performed in two distinct cases: when the helicities of the produced fermions are equal and when these quantities are opposite as sign. The basic results were related to the fact that in the Minkowski limit the probability density is vanishing \cite{10,19,20,21}. Another consequence was related to the helicity conservation in this process and the results prove that, in the case when helicity is not conserved $\sigma=\sigma'$ the density probability is vanishing in the limit $k\rightarrow 0$, while for the case of helicity conservation $\sigma=-\sigma'$, the density probability is finite in the same limit. For obtaining the total probability an integration after the final momenta $p\,,p'$ is needed, and we expect as a general rule, to preserve the same behaviour in terms of parameter $k$, in the total probability.

The amplitude equation (\ref{an1}), can be rewritten using Eq.(\ref{f}) and equation (\ref{1}) from Appendix, by evaluating the quantity $f_{k}^{*}\left(\frac{p'}{p}\right)-sgn(\sigma\sigma')f_{k}\left(\frac{p'}{p}\right)$.
The first step is to write the momenta ratio at imaginary powers in the form:
\begin{equation}\label{eul}
\left(\frac{p'}{p}\right)^{ik}= e^{ik\ln\left(\frac{p'}{p}\right)}= \cos \left(k\ln\left(\frac{p'}{p}\right)\right)+ i\sin \left(k\ln\left(\frac{p'}{p}\right)\right).
\end{equation}
The second step is to compute the addition and difference of the $f_k\left(\frac{p'}{p}\right)$ functions, and in the case $\sigma=\sigma'$ this gives:
\begin{equation}
f_{k}^{*}\left(\frac{p'}{p}\right)- f_{k}\left(\frac{p'}{p}\right)= -\frac{2i\pi \sin \left(k\ln\left(\frac{p'}{p}\right)\right)}{\cosh(\pi k)\left(1-\frac{p'}{p}\right)},
\end{equation}
while in the case $\sigma=-\sigma'$ the result is:
\begin{equation}
f_{k}^{*}\left(\frac{p'}{p}\right)+ f_{k}\left(\frac{p'}{p}\right)= \frac{2\pi \cos \left(k\ln\left(\frac{p'}{p}\right)\right)}{\cosh(\pi k)\left(1+\frac{p'}{p}\right)}.
\end{equation}
Further we introduce the following notation:
\begin{equation}
M_{\sigma\sigma'}=\xi_{\sigma}^{+}(\vec{p}\,)[\vec{\sigma}\cdot(\vec{\mathcal{M}}\times(\vec{p}+\vec{p}\,'))]\eta_{\sigma'}(\vec{p}\,'),
\end{equation}
then the transition amplitude equation becomes:
\begin{eqnarray}\label{af}
\mathcal{A}_{e^{-}e^{+}}= \frac{-ie\,M_{\sigma\sigma'}}{2\pi^{2}|\vec{p}+\vec{p}\,'|^{2}\cosh(\pi k)}\frac{\theta(p-p')}{p}
\times\left\{
\begin{array}{cll}
\frac{-i \sin \left(k\ln\left(\frac{p'}{p}\right)\right)}{\left(1-\frac{p'}{p}\right)}&{\rm for}&\sigma=\sigma'\\
\frac{ \cos \left(k\ln\left(\frac{p'}{p}\right)\right)}{\left(1+\frac{p'}{p}\right)}&{\rm for}&\sigma=-\sigma'.
\end{array}\right.
\end{eqnarray}
The presence of the trigonometric functions of argument proportional with $k$ explain the behaviour obtained in \cite{10} for the probability density in the limit $k\rightarrow0$, since the sine function of zero argument is vanishing and the cosine function is finite for a null argument.
The probability density is obtained by summing after the final polarizations $\sigma,\sigma'$ the square modulus of the transition amplitude. Therefore, using equation (\ref{af}) we obtain:
\begin{eqnarray}\label{dp1}
\mathcal{P}_{e^-e^+}=\frac{1}{2}\sum_{\sigma \sigma'}|M_{\sigma\sigma'}|^2\frac{e^2}{4\pi^{4}|\vec{p}+\vec{p}\,'|^{4}\cosh^2(\pi k)}\frac{1}{p^2}
\times\left\{
\begin{array}{cll}
\frac{ \sin^2 \left(k\ln\left(\frac{p'}{p}\right)\right)}{\left(1-\frac{p'}{p}\right)^2}&{\rm for}&\sigma=\sigma'\\
\frac{ \cos^2 \left(k\ln\left(\frac{p'}{p}\right)\right)}{\left(1+\frac{p'}{p}\right)^2}&{\rm for}&\sigma=-\sigma'.
\end{array}\right.
\end{eqnarray}
Equations (\ref{af}) and (\ref{dp1}) prove that our results for the probability density obtained earlier \cite{10}, can be brought in a form that can facilitate the computations for the total probability.

\subsection{The total probability}
The total probability is obtained by integrating the equation (\ref{dp1}) after the final momenta. First we specify that for doing the summation in the equation (\ref{dp1}) we work with the spherical coordinates for the momenta such that $\vec{p}=(p,\alpha,\beta)$, and
${\vec{p}\,}\,'=(p\,',\gamma,\varphi)$, where $\alpha,\, \gamma\in(0,\pi);\,\beta,\, \varphi\in(0,2\pi)$ \cite{10}. Second, we recall the result obtained in \cite{10} for the probability density after summation. Then fixing $\alpha=\gamma=\pi/2$, we obtain the situation when the momenta components are in the $(1,2)$ plane and the results will be obtained in terms of the azimuthal angles $\beta,\varphi$ \cite{10}. This choice is justified by the fact that the probability of fermion production is maximum when the momenta vectors of the fermions are perpendicular on the direction of the magnetic dipole \cite{10}. Collecting all the results and introducing the fine structure constant $\alpha=\frac{e^2}{4\pi}$, the probability density becomes:
\begin{eqnarray}\label{dp}
\mathcal{P}_{e^-e^+}=\frac{\alpha\mathcal{M}^2}{2\pi^{3}(p^2+p\,'^2+2pp\,'\cos(\beta-\varphi))^2\cosh^2(\pi k)}\frac{1}{p^2}\nonumber\\
\times\left\{
\begin{array}{cll}
\frac{ \sin^2 \left(k\ln\left(\frac{p'}{p}\right)\right)}{\left(1-\frac{p'}{p}\right)^2}(p+p\,')^2(1+\cos(\beta-\varphi))&{\rm for}&\sigma=\sigma'\\
\frac{ \cos^2 \left(k\ln\left(\frac{p'}{p}\right)\right)}{\left(1+\frac{p'}{p}\right)^2}(p-p\,')^2(1-\cos(\beta-\varphi))&{\rm for}&\sigma=-\sigma'.
\end{array}\right.
\end{eqnarray}
The total probability can be calculated using the above equation as:
\begin{equation}\label{pt}
\mathcal{P}^{tot}_{e^-e^+}=\int \mathcal{P}_{e^-e^+}\,d^3p\,d^3p\,'.
\end{equation}
We specify that our computation for the total probability will be done by fixing the angle between momenta vectors $\beta-\varphi$. This is equivalently to compute the total probability for fixed directions on which the particles are emitted, and this requires to solve only the momenta modulus integrals.

\subsection{Helicity nonconserving case $\sigma=\sigma'$}
The general form of the total probability equation for $\sigma=\sigma'$, is obtained using equations (\ref{dp}) and (\ref{pt}):

\begin{equation}
\mathcal{P}^{tot}_{\sigma=\sigma'}=\frac{\alpha\mathcal{M}^2(1+\cos(\beta-\varphi))}{2\pi^{3}\cosh^2(\pi k)}\int_{0}^{\infty} \,dp\,'p'^2\,\int_{0}^{\infty}dp\frac{\sin^2 \left(k\ln\left(\frac{p'}{p}\right)\right)(p+p\,')^2}{\left(1-\frac{p'}{p}\right)^2(p^2+p\,'^2+2pp\,'\cos(\beta-\varphi))^2}.
\end{equation}
The integral after fermion momentum $p$ is not directly solvable and for that reason we can approximate the integrand. In the second integral after the anti-fermion momentum $p'$, we apply a cutoff of the upper limit up to $P'/\omega$, which is an adimensional factor. Since the argument of the sine function contains the parameter $k$ and we prove in \cite{10} that the probability density has nonvanishing values only for small $k$, we use the Taylor expansion of both sine function and natural logarithm and restrict to the first term in our further considerations:
\begin{eqnarray}\label{ty}
&&\sin \left(k\ln\left(\frac{p'}{p}\right)\right)\simeq k\ln\left(\frac{p'}{p}\right),\nonumber\\
&&\ln\left(\frac{p'}{p}\right)\simeq\frac{p'}{p}-1.
\end{eqnarray}
We must specify that our approximations are good when we use the Taylor expansion of our trigonometric functions and logarithm as long as the momenta ratio are taken between $ \frac{p'}{p}\in(0.2, 0.9)$, and this can be checked by numerical calculations .
The equation for the total probability in terms of angle $\beta-\varphi$ can be rewritten by using (\ref{ty}) in the form:
\begin{equation}
\mathcal{P}^{tot}_{\sigma=\sigma'}=\frac{\alpha\mathcal{M}^2k^2(1+\cos(\beta-\varphi))}{2\pi^{3}\cosh^2(\pi k)}\int_{0}^{P'/\omega} \,dp\,'p'^2\,\int_{0}^{\infty}dp\frac{(p+p\,')^2}{(p^2+p\,'^2+2pp\,'\cos(\beta-\varphi))^2}.
\end{equation}
The general form of the integrals that help us to compute the total probability are presented in the Appendix. Here we present the final results for different values of the angle between momenta vectors $\beta-\varphi$:
\begin{eqnarray}\label{ss}
\mathcal{P}^{tot}_{\sigma=\sigma'}=\frac{\alpha\mathcal{M}^2k^2}{2\pi^{3}\cosh^2(\pi k)}\left(\frac{P'}{\omega}\right)^2\times\left\{
\begin{array}{cll}
1\,&{\rm for}&\beta-\varphi=0\\
\left(\frac{1}{2}+\frac{\sqrt{3}}{4}\right)\left(\frac{4\pi}{3}-\frac{2\pi\sqrt{3}}{3} -\frac{6\sqrt{3}}{3}+4\right)\,&{\rm for}&\beta-\varphi=\pi/6\\
\frac{3}{4}\left(\frac{4\pi\sqrt{3}}{27}+ \frac{18}{27}\right)&{\rm for}&\beta-\varphi=\pi/3\\
\frac{1}{4}\left(\frac{8\pi\sqrt{3}}{27}+ \frac{18}{27}\right)\,&{\rm for}&\beta-\varphi=2\pi/3\\
\left(\frac{1}{2}-\frac{\sqrt{3}}{4}\right)\left(\frac{20\pi}{3}+\frac{10\pi\sqrt{3}}{3} +\frac{6\sqrt{3}}{3}+4\right)\,&{\rm for}&\beta-\varphi=5\pi/6.
\end{array}\right.
\end{eqnarray}
When we set $\beta-\varphi=\pi$, the total probability in the case $\sigma=\sigma'$, is vanishing due to the factor $1+\cos(\beta-\varphi)$.

\subsection{Helicity conserving case $\sigma=-\sigma'$}

In the case $\sigma= -\sigma'$ the total probability equation obtained from equations (\ref{dp}) and (\ref{pt})reads:
\begin{equation}
\mathcal{P}^{tot}_{\sigma=-\sigma'}=\frac{\alpha\mathcal{M}^2(1-\cos(\beta-\varphi))}{2\pi^{3}\cosh^2(\pi k)}\int_{0}^{\infty} \,dp\,'p'^2\,\int_{0}^{\infty}dp\frac{\cos^2 \left(k\ln\left(\frac{p'}{p}\right)\right)(p-p\,')^2}{\left(1+\frac{p'}{p}\right)^2(p^2+p\,'^2+2pp\,'\cos(\beta-\varphi))^2}.
\end{equation}
We use further the Taylor expansion of cosine and retain only the first term :
\begin{equation}
\cos \left(k\ln\left(\frac{p'}{p}\right)\right)\simeq 1,
\end{equation}
and finally obtain the total probability equation in the form:
\begin{eqnarray}
\mathcal{P}^{tot}_{\sigma=-\sigma'}=\frac{\alpha\mathcal{M}^2(1-\cos(\beta-\varphi))}{2\pi^{3}\cosh^2(\pi k)}\int_{0}^{P'/\omega} \,dp\,'p'^2\,\int_{0}^{\infty}dp\frac{p^2(p-p\,')^2}{\left(p+p'\right)^2(p^2+p\,'^2+2pp\,'\cos(\beta-\varphi))^2}.\nonumber\\
\end{eqnarray}
For $\beta-\varphi=0$ the total probability is vanishing. The final result for different values of $\beta-\varphi$ is given bellow and the integrals that help us to compute the total probability are given in Appendix.
\begin{eqnarray}\label{chp}
\mathcal{P}^{tot}_{\sigma=-\sigma'}=\frac{\alpha\mathcal{M}^2}{2\pi^{3}\cosh^2(\pi k)}\left(\frac{P'}{\omega}\right)^2D\left[A + B\left(\ln\left(\frac{P'}{\omega}\right)-\frac{1}{2}\right)+ C\left(\ln\left(\frac{P'}{\omega}\right)^2-1\right)\right],
\end{eqnarray}
where the coefficients $A,B,C,D$ are numerical factors that depend on the given values for the angle between momenta vectors, $\beta-\varphi$. We observe the dependence on the logarithmic factor $\ln\left(\frac{P'}{\omega}\right)$, which proves that in the limit of large momenta the divergences are logarithmic.

The values of the coefficients $A,B,C,D$ can be found bellow, and for $\beta-\varphi=\pi/6$ we obtain:
\begin{eqnarray}\label{c1}
&&A= \frac{2\pi+2\pi\sqrt{3}}{-7+4\sqrt{3}}+ \frac{20\pi+27\sqrt{3}-48-12\pi\sqrt{3}}{168\sqrt{3}-291},\,\,\,\,\,
B= \frac{48\sqrt{3}-84}{168\sqrt{3}-291},\nonumber\\
&&C= \frac{42-24\sqrt{3}}{168\sqrt{3}-291},\,\,\,\,\,D=\frac{1}{2}+\frac{\sqrt{3}}{4}.
\end{eqnarray}
When $\beta-\varphi=\pi/3$ one finds that:
\begin{equation}\label{c2}
A= -\frac{8\pi\sqrt{3}}{9}+ 5,\,\,\,B= 4,\,\,\,C= -2,D=\frac{1}{4},
\end{equation}
while for $\beta-\varphi=2\pi/3$:
\begin{equation}\label{c3}
A= \frac{1}{3},\,\,\,B= \frac{4}{9},\,\,\,C= -\frac{2}{9},D=\frac{3}{4}.
\end{equation}

\subsection{Limit cases}
At the end of this section we will discuss the limit $\frac{m}{\omega}=0$, which corresponds to the situation when the expansion parameter is much more larger than the fermion mass. In the helicity nonconserving case the limit $\frac{m}{\omega}=0$ cancel the total probability, as can be seen from equation (\ref{ss}), while in the case of helicity conservation we obtain a finite quantity:
\begin{eqnarray}
&&\mathcal{P}^{tot}_{\sigma=\sigma'}|_{(k=0)}=0 ,\nonumber\\
&&\mathcal{P}^{tot}_{\sigma=-\sigma'}|_{(k=0)}=\frac{\alpha\mathcal{M}^2}{2\pi^{3}}\left(\frac{P'}{\omega}\right)^2D\left[A + B\left(\ln\left(\frac{P'}{\omega}\right)-\frac{1}{2}\right)+ C\left(\ln\left(\frac{P'}{\omega}\right)^2-1\right)\right],\,\,\,\,\,
\end{eqnarray}
with the coefficients $A,B,C,D$ given above for fixed values of $\beta-\varphi$.

The second limit of the analytical results is related to the case when the mass of the fermion is much more larger than the expansion parameter $m>>\omega$. In this limit our equations for total probability become:
\begin{eqnarray}\label{ssl}
\mathcal{P}^{tot}_{\sigma=\sigma'}|_{(k>>1)}=\frac{\alpha\mathcal{M}^2\left(\frac{m}{\omega}\right)^2e^{-2\pi \frac{m}{\omega}}}{2\pi^{3}}\left(\frac{P'}{\omega}\right)^2\times\left\{
\begin{array}{cll}
1\,&{\rm for}&\beta-\varphi=0\\
\left(\frac{1}{2}+\frac{\sqrt{3}}{4}\right)\left(\frac{4\pi}{3}-\frac{2\pi\sqrt{3}}{3} -\frac{6\sqrt{3}}{3}+4\right)\,&{\rm for}&\beta-\varphi=\pi/6\\
\frac{3}{4}\left(\frac{4\pi\sqrt{3}}{27}+ \frac{18}{27}\right)\,&{\rm for}&\beta-\varphi=\pi/3\\
\frac{1}{4}\left(\frac{8\pi\sqrt{3}}{27}+ \frac{18}{27}\right)\,&{\rm for}&\beta-\varphi=2\pi/3\\
\left(\frac{1}{2}-\frac{\sqrt{3}}{4}\right)\left(\frac{20\pi}{3}+\frac{10\pi\sqrt{3}}{3} +\frac{6\sqrt{3}}{3}+4\right)\,&{\rm for}&\beta-\varphi=5\pi/6.
\end{array}\right.
\end{eqnarray}
\begin{eqnarray}\label{chpl}
\mathcal{P}^{tot}_{\sigma=-\sigma'}|_{(k>>1)}=\frac{\alpha\mathcal{M}^2e^{-2\pi \frac{m}{\omega}}}{2\pi^{3}}\left(\frac{P'}{\omega}\right)^2D\left[A + B\left(\ln\left(\frac{P'}{\omega}\right)-\frac{1}{2}\right)+ C\left(\ln\left(\frac{P'}{\omega}\right)^2-1\right)\right],\,\,\,\,\,\,\,
\end{eqnarray}
where the coefficients $A,B,C,D$ are given in equations (\ref{c1})-(\ref{c3}). The total probability in this limit become small since we have the dependence on the factor $e^{-2\pi \frac{m}{\omega}}$, and we conclude that for large fermion mass comparatively with the expansion parameter the probability becomes negligible. The exponent $e^{-2\pi \frac{m}{\omega}}$ was also found in \cite{31,32,33}, as the factor that is proportional with the density number of particles produced in external fields, and the results were obtained using nonperturbative methods. We also specify that there is a extended literature on the nonperturbative treatment on the phenomenon of particle production \cite{17,24,25,26,27,31,32,33}, which can be observed by using local detectors, while in our case the global apparatus can record particle generation only in the presence of electromagnetic interactions \cite{7}.

\subsection{Graphical analysis and discussion}
The topic of this section will be the graphical analysis of the total probability in terms of parameter $m/\omega$. In all graphs we set the magnetic dipole modulus to $\mathcal{M}=1$ and $\frac{P'}{\omega}=1$ in the case $\sigma=\sigma'$. Since for $\sigma=-\sigma'$ the total probability is proportional with natural logarithm $\ln\left(\frac{P'}{\omega}\right)$, then for avoiding zero values of logarithm we take $\frac{P'}{\omega}=0.9$, with the observation that the graphs can be also done for $\frac{P'}{\omega}=1$. The fine structure constant is dimensionless and we use its value $\alpha=\frac{1}{137}$ in all graphs.

The adimensional parameter $\frac{P'}{\omega}$ has an important role in our equations for the total probability and its value can be estimated by recalling the Compton wavelength for the electron. The momenta can be  written in terms of wavelength as, $p=h/\lambda$. Then the upper limit of the integral $\frac{P'}{\omega}$ will be:
\begin{equation}
\frac{p}{\omega}=\frac{h}{\omega\lambda_{C}}=\frac{m\,c}{\omega},\, \lambda_{C}=\frac{h}{m c},
\end{equation}
where $\lambda_{C}$ is the Compton wavelength associated to electron and it is proportional with $m^{-1}$. Then if we use natural units such that $c=1$, $h=1$ we conclude that the quantity $\frac{P'}{\omega}\leq\frac{m}{\omega}$. Then in our further graphical analysis $\frac{P'}{\omega}$ will be taken smaller or equal with one, since the graphical analysis from \cite{10}, proves that the probability density has nonvanishing values only in the interval $\frac{m}{\omega}\in[0,1]$. The graphs shows the total probability dependence in terms of parameter $k$.

\begin{figure}[h!t]
\includegraphics[scale=0.45]{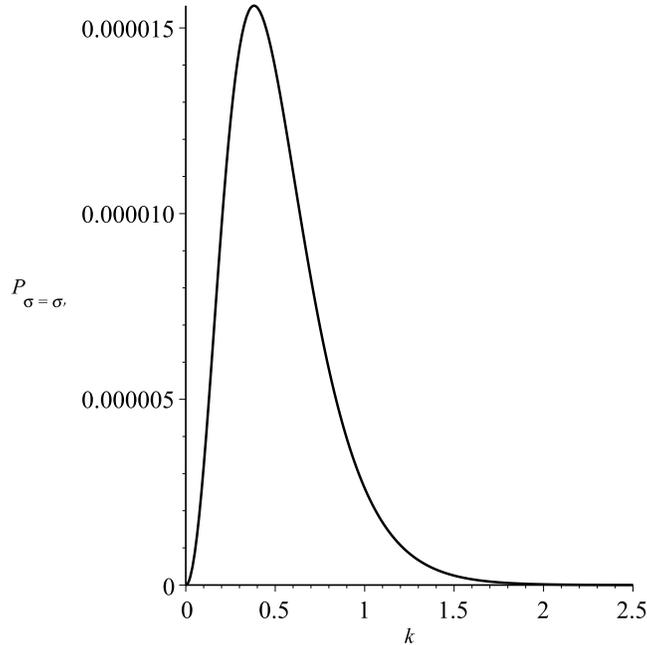}
\caption{Total probability as a function of parameter $k$ for $\beta-\varphi=\frac{\pi}{6}$, in the case $\sigma=\sigma'$.}
\label{f1}
\end{figure}

\begin{figure}[h!t]
\includegraphics[scale=0.45]{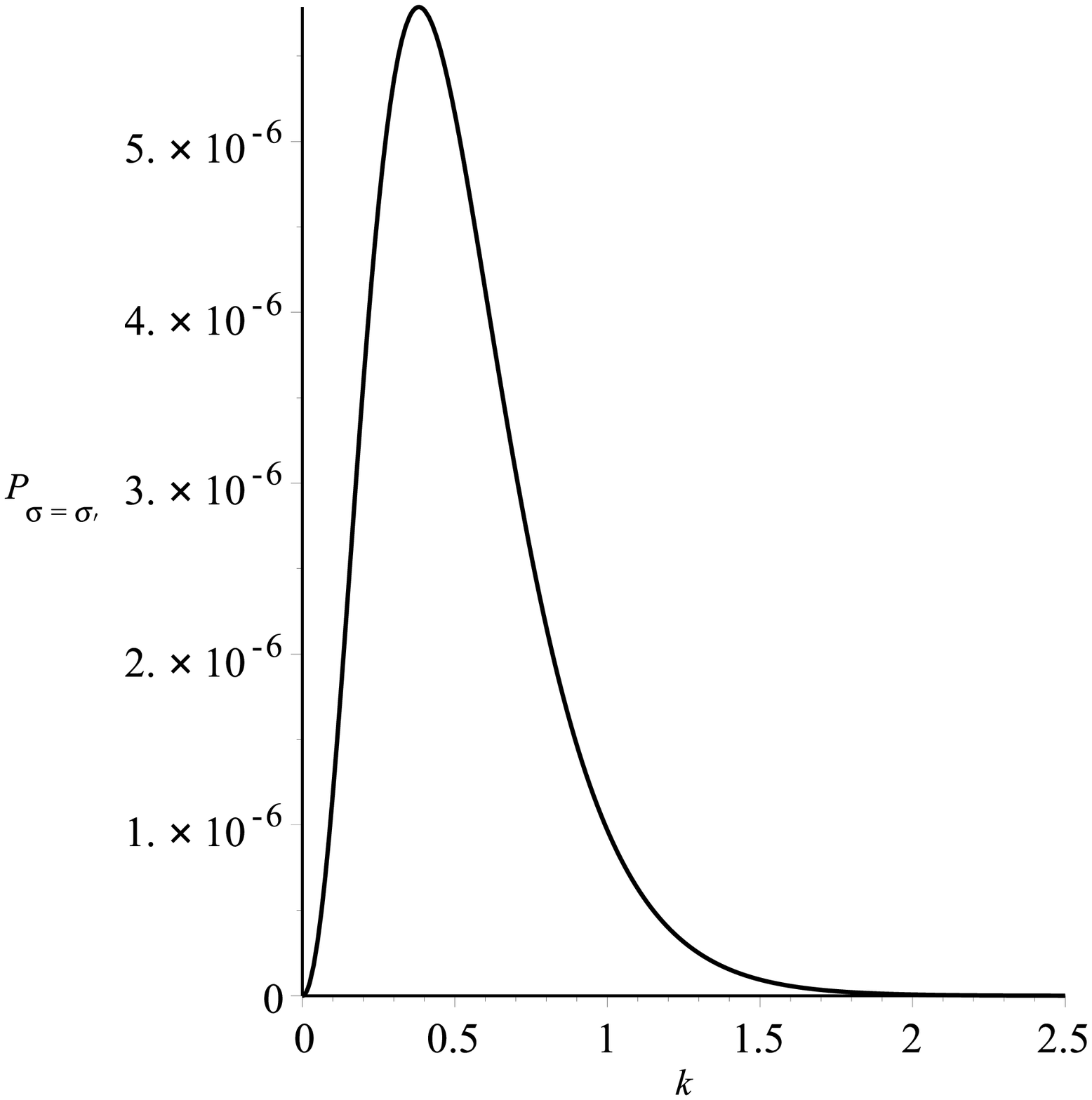}
\caption{Total probability as a function of parameter $k$ for $\beta-\varphi=\frac{\pi}{3}$, in the case $\sigma=\sigma'$.}
\label{f2}
\end{figure}

\begin{figure}[h!t]
\includegraphics[scale=0.45]{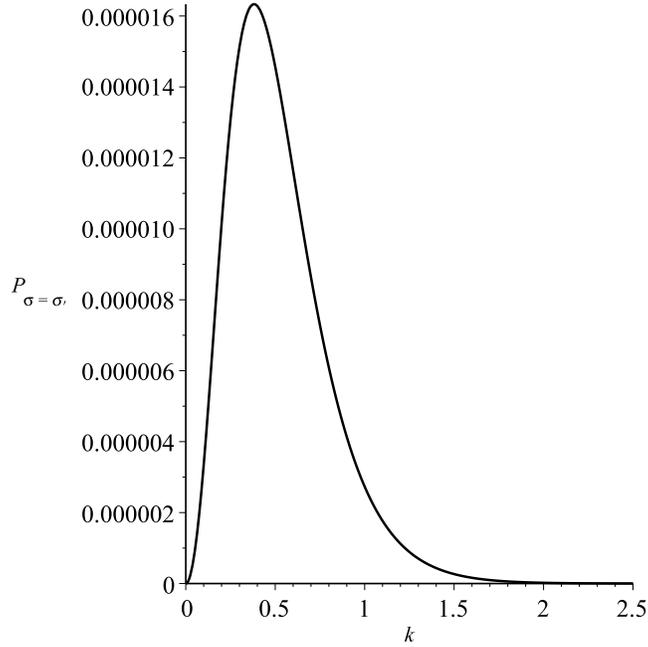}
\caption{Total probability as a function of parameter $k$ for $\beta-\varphi=\frac{5 \pi}{6}$, in the case $\sigma=\sigma'$.}
\label{f3}
\end{figure}

The Figs.(\ref{f1})-(\ref{f3}) proves that the total probability in the case $\sigma=\sigma'$, is nonvanishing only for small values of the parameter $k$. For $k=0$, the total probability is vanishing and this result confirms the results obtained in \cite{10}, where the probability density is also vanishing for $k=0$. Another result is related to the fact that we prove the total probability dependence on the angle between the momenta of the produced particles.

\begin{figure}[h!t]
\includegraphics[scale=0.45]{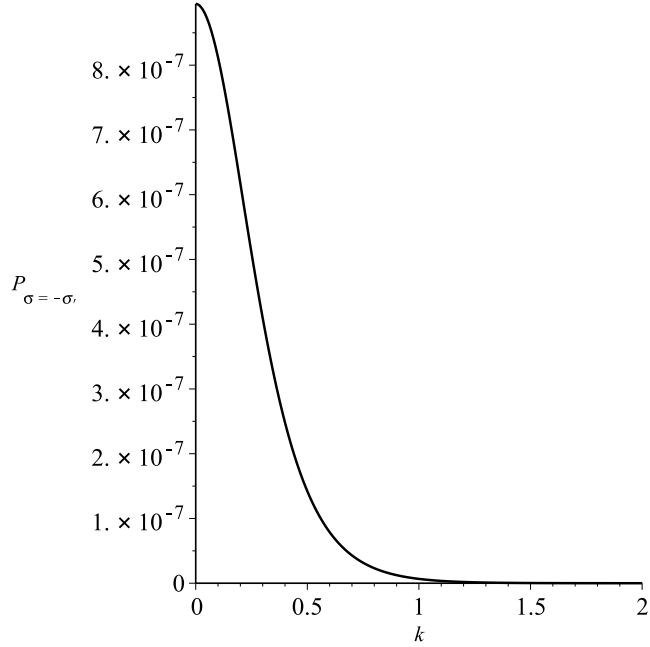}
\caption{Total probability as a function of parameter $k$ for $\beta-\varphi=\frac{\pi}{6}$, in the case $\sigma=\sigma'$.}
\label{f4}
\end{figure}

\begin{figure}[h!t]
\includegraphics[scale=0.45]{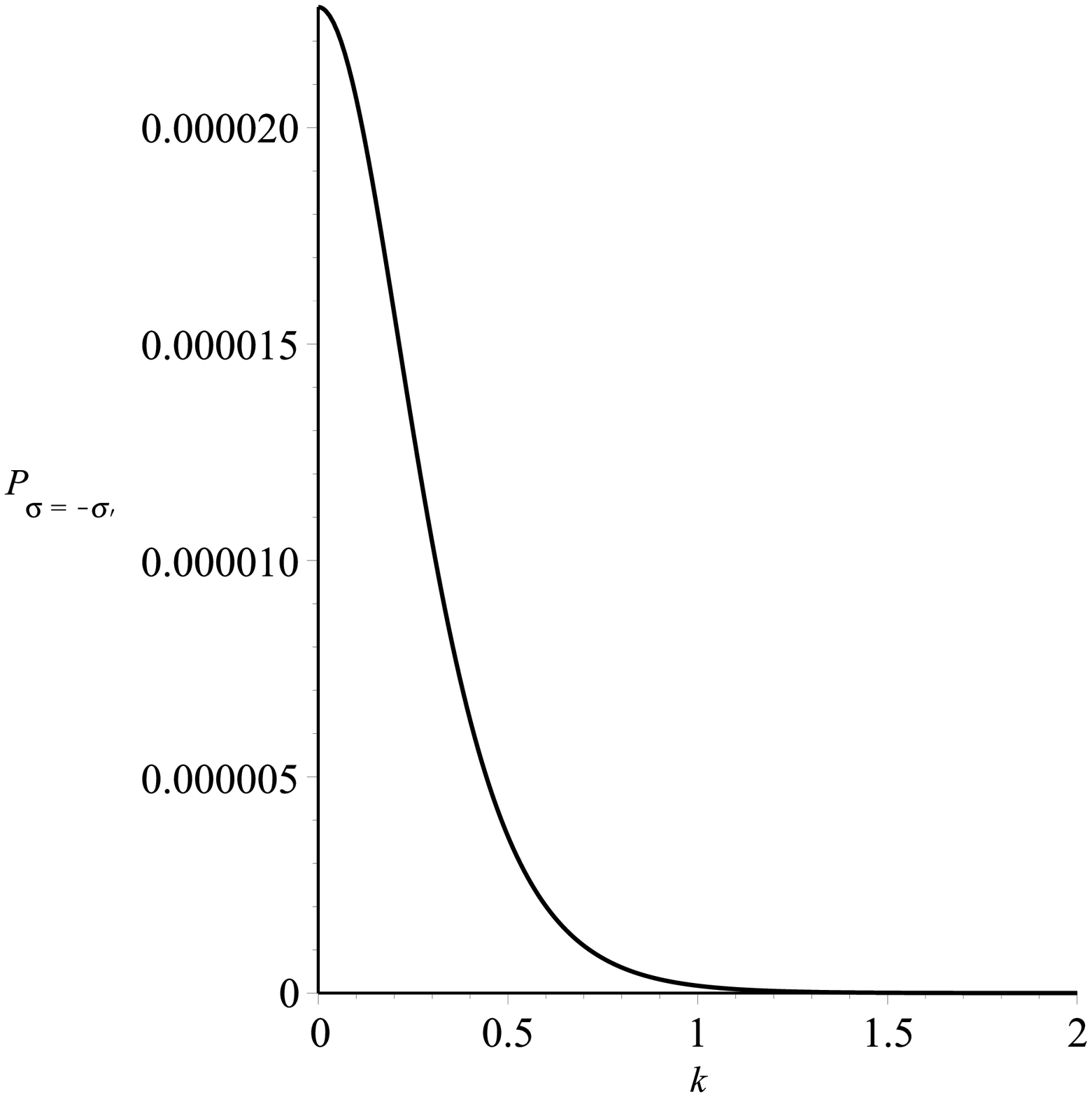}
\caption{Total probability as a function of parameter $k$ for $\beta-\varphi=\frac{\pi}{3}$, in the case $\sigma=-\sigma'$.}
\label{f5}
\end{figure}

\begin{figure}[h!t]
\includegraphics[scale=0.45]{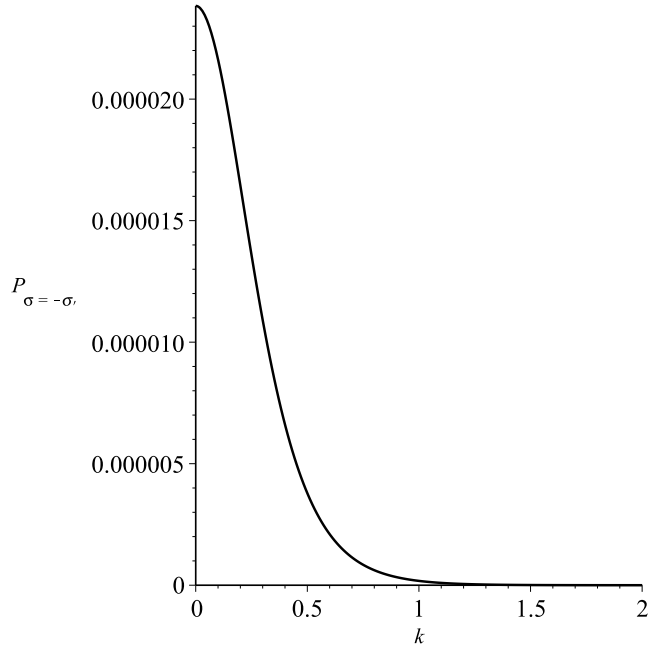}
\caption{Total probability as a function of parameter $k$ for $\beta-\varphi=\frac{2 \pi}{3}$, in the case $\sigma=-\sigma'$.}
\label{f6}
\end{figure}

In the case of helicity conservation $\sigma=-\sigma'$ (Figs.(\ref{f4})-(\ref{f6})) we obtain that, the total probability equation contains the result $k=0$ (see equation (\ref{chp})), since in the limit $k\rightarrow0$ this quantity is no longer vanishing. In fact this result contains the case when the expansion parameter is much more larger than the fermion mass $\omega>>m$. The variation with the angle between momenta of the fermions $\beta-\varphi$, shows that the total probability is increasing when $\beta-\varphi$ become larger.

Comparing the total probabilities obtained above we observe that the processes that conserve the helicity ($\sigma=-\sigma'$), are favoured under the space expansion, since in this situation the total probability increases as the angle between fermions momenta approaches $\pi$  .

Another result obtained in Figs.(\ref{f1})-(\ref{f6}) is related to the fact that the total probability becomes negligible for $m>\omega$ and is vanishing in the Minkowski limit, where the generation of fermion pairs in dipole magnetic field is forbidden as a perturbative phenomenon by the energy and momentum conservation.

\section{Fermion production in Coulomb Field}
In this section the total probability for fermion production in Coulomb field on de Sitter geometry will be analysed. The amplitude of pair production obtained in \cite{13} will be written in a compact form by using the Bessel $K$ functions:
\begin{eqnarray}\label{in}
\mathcal{A}_{e^-e^+}=
\frac{e^{2} Z\sqrt{pp\,'}}{2\pi^3|\vec{p}+\vec{p}\,'|^{2}}\left[-sgn(\lambda\,')\int_0^{\infty} dz
zK_{\nu_{+}}(ip z)K_{\nu_{-}}(ip\,'z)\right.\nonumber\\
\left.+sgn(\lambda)\int_0^{\infty} dz
zK_{\nu_{-}}(ip z)K_{\nu_{+}}(ip\,' z)\right]\xi^{+}_{\lambda}(\vec{p}\,)\eta_{\lambda'}(\vec{p}\,\,'),
\end{eqnarray}
The integrals that help us to establish the final result can be found in \cite{16,22}.
Then the amplitude give for the case $p>p'$ :
\begin{equation}\label{a1}
\mathcal{A}_{e^{-}e^{+}}= -\frac{ie^2Z}{8\pi^{2}|\vec{p}+\vec{p}\,'|^{2}}\frac{\theta(p-p')}{p}\left[-sgn(\sigma')f_{k}\left(\frac{p'}{p}\right)+sgn(\sigma)f^{*}_{k}\left(\frac{p'}{p}\right)\right]
\xi_{\sigma}^{+}(\vec{p}\,)\eta_{\sigma'}(\vec{p}\,'),
\end{equation}
where we define the functions:
\begin{equation}
f_{k}\left(\frac{p'}{p}\right)=\left(\frac{p'}{p}\right)^{1-ik}\Gamma(1-ik)\Gamma(1+ik)_{2}F_{1}\left(1-ik;\frac{3}{2};2;1-\left(\frac{p\,'}{p}\right)^{2}\right).
\end{equation}
The above function can be rewritten using $\Gamma(1-ik)\Gamma(1+ik)=\frac{\pi k}{\sinh(\pi k)}$. The hypergeometric function cannot be reduced to a polynomial factor as in the case of fermion generation in magnetic field, and the momenta integrals with two hypergeometric functions and a polynom at the denominator are not known in literature. This situation suggests for looking an approximation for solving the momenta integrals to obtain the total probability.We will look for a suitable function to approximate the hypergeometric function $_{2}F_{1}\left(1-ik;\frac{3}{2};2;1-\left(\frac{p\,'}{p}\right)^{2}\right)$, such that the dependence on momenta and the factor $k$ to be preserved in the new function
\begin{equation}\label{apx}
_{2}F_{1}\left(1-ik;\frac{3}{2};2;1-\left(\frac{p\,'}{p}\right)^{2}\right)\simeq k^{-1/3}\left(\frac{p\,'}{p}\right)^{-2+2ik}.
\end{equation}
A graphical analysis proves that the hypergeometric function is well approximated by the function given above and that the approximate function is convergent for large $k$ in both real part and imaginary part. The approximation is well justified for $\frac{p\,'}{p}\in(0.4,0.99)$ or in other words when the ratio of the momenta is close to unity.
\begin{figure}[h!t]
\includegraphics[scale=0.45]{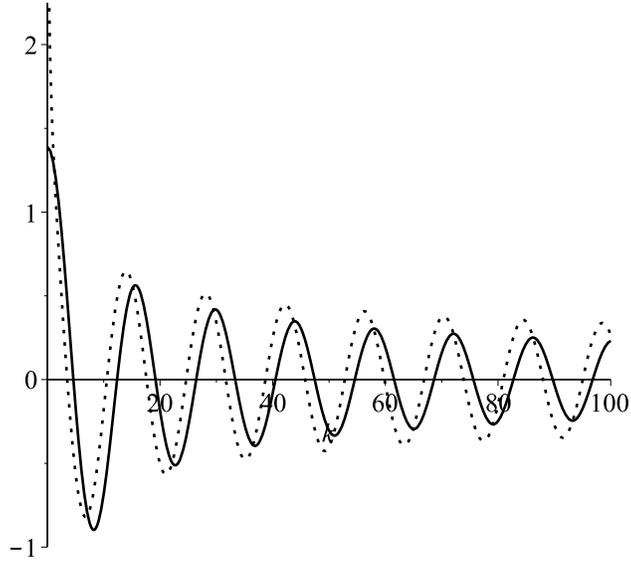}
\caption{Real part of hypergeometric function (solid line) and approximate function (point line) in terms of parameter $k$ for $\frac{p\,'}{p}=0.8$.}
\label{f7}
\end{figure}

\begin{figure}[h!t]
\includegraphics[scale=0.45]{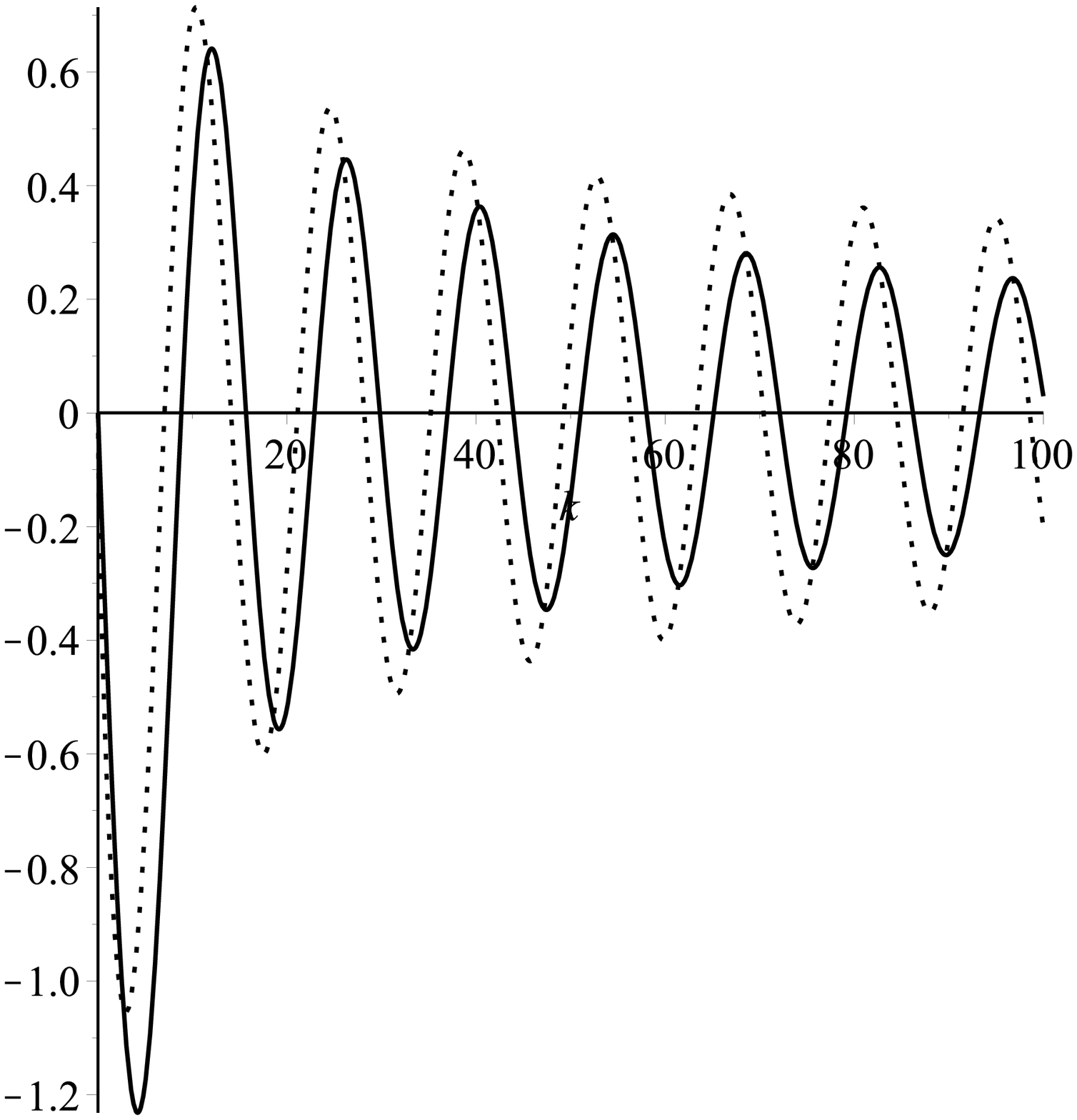}
\caption{Imaginary part of hypergeometric function (solid line) and approximate function (point line) in terms of parameter $k$ for $\frac{p\,'}{p}=0.8$.}
\label{f8}
\end{figure}

\begin{figure}[h!t]
\includegraphics[scale=0.45]{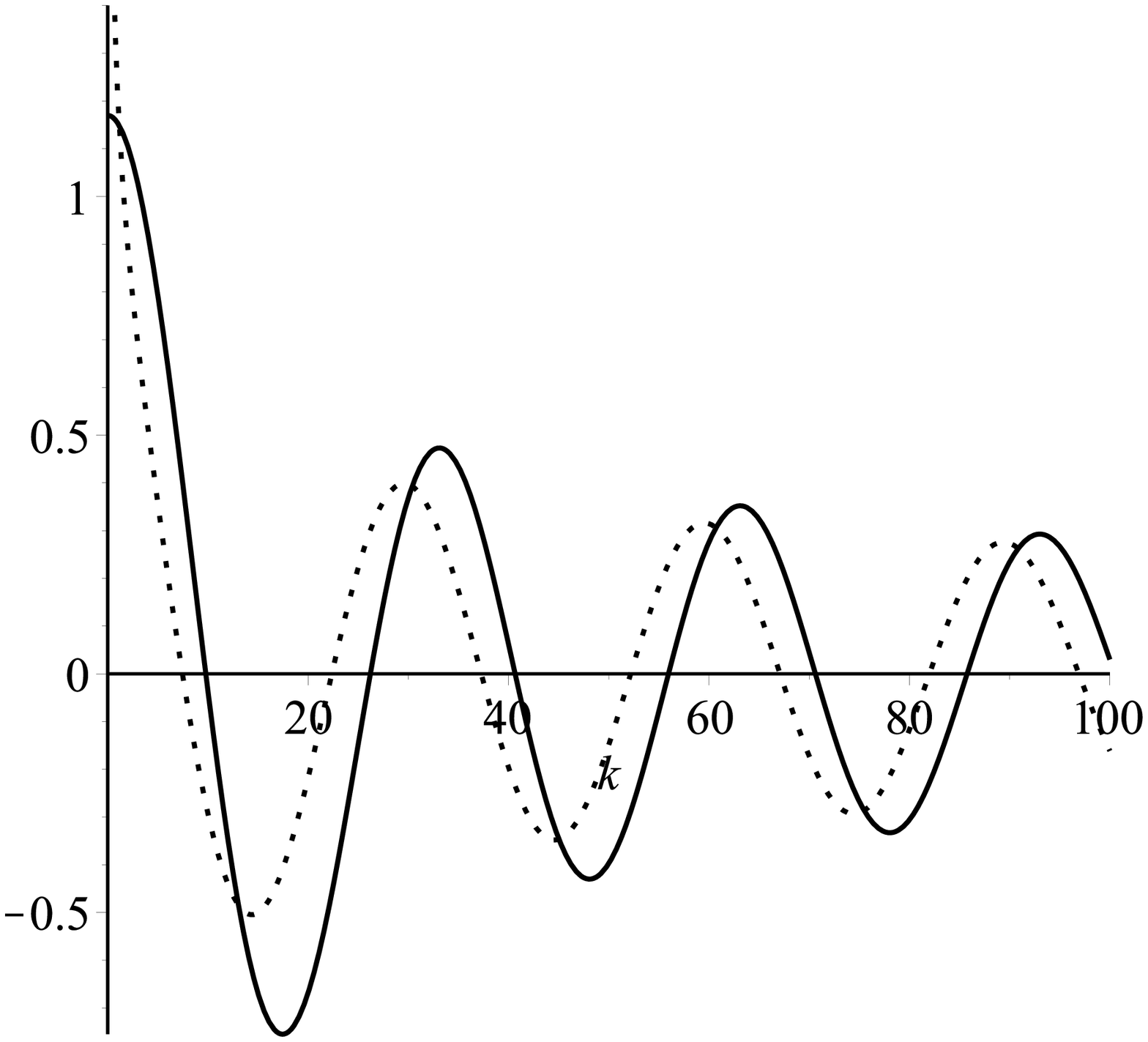}
\caption{Real part of hypergeometric function (solid line) and approximate function (point line) in terms of parameter $k$ for $\frac{p\,'}{p}=0.9$.}
\label{f9}
\end{figure}

\begin{figure}[h!t]
\includegraphics[scale=0.45]{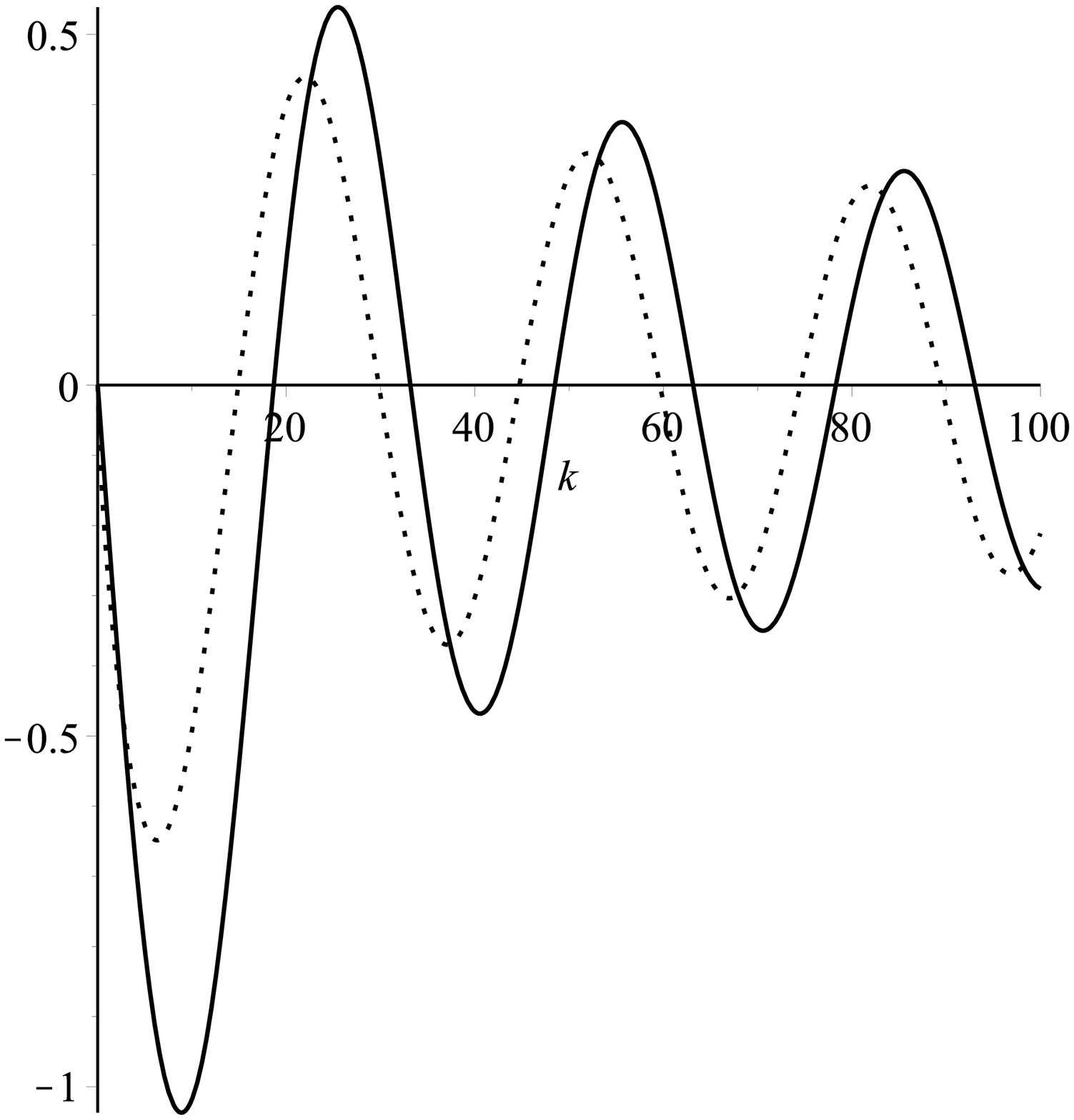}
\caption{Imaginary part of hypergeometric function (solid line) and approximate function (point line) in terms of parameter $k$ for $\frac{p\,'}{p}=0.9$.}
\label{f10}
\end{figure}

Then the form of function $f_{k}\left(\frac{p'}{p}\right)$ becomes if we use equation (\ref{apx}):
\begin{equation}
f_{k}\left(\frac{p'}{p}\right)=\frac{\pi k^{2/3}}{\sinh(\pi k)}\left(\frac{p'}{p}\right)^{-1+ik}.
\end{equation}
Now we use equation (\ref{eul}) to express the momenta ratio at imaginary powers and compute the form of the amplitudes for $\sigma=\sigma'$ and $\sigma=-\sigma'$ :
\begin{eqnarray}\label{afcy}
\mathcal{A}_{e^{-}e^{+}}= \frac{e^2Zk^{2/3}}{4\pi|\vec{p}+\vec{p}\,'|^{2}\sinh(\pi k)}\left(\frac{p'}{p}\right)^{-1}\frac{\theta(p-p')}{p}\xi_{\sigma}^{+}(\vec{p}\,)\eta_{\sigma'}(\vec{p}\,')
\times\left\{
\begin{array}{cll}
-i \sin \left(k\ln\left(\frac{p'}{p}\right)\right)&{\rm for}&\sigma=\sigma'\\
\cos \left(k\ln\left(\frac{p'}{p}\right)\right)&{\rm for}&\sigma=-\sigma'.
\end{array}\right.
\end{eqnarray}
The probability of transition is obtained by summing after the final polarizations $\sigma,\sigma'$ the square modulus of the amplitude:
\begin{eqnarray}\label{afcpy}
\mathcal{P}_{e^{-}e^{+}}= \frac{1}{2}\sum_{\sigma\sigma'}\frac{e^4Z^2k^{4/3}}{16\pi^2|\vec{p}+\vec{p}\,'|^{4}\sinh^2(\pi k)}\frac{1}{p'\,^2}|\xi_{\sigma}^{+}(\vec{p}\,)\eta_{\sigma'}(\vec{p}\,')|^2
\times\left\{
\begin{array}{cll}
\sin^2 \left(k\ln\left(\frac{p'}{p}\right)\right)&{\rm for}&\sigma=\sigma'\\
\cos^2 \left(k\ln\left(\frac{p'}{p}\right)\right)&{\rm for}&\sigma=-\sigma'.
\end{array}\right.
\end{eqnarray}

\subsection{Total probability for fermion production in Coulomb field}
Integration of the equation (\ref{afcpy}) after the final momenta gives the total probability of the process. As in the previous section we take the spherical coordinates for the momenta such that $\vec{p}=(p,\alpha,\beta)$ and
${\vec{p}\,}\,'=(p\,',\gamma,\varphi)$, where $\alpha,\, \gamma\in(0,\pi);\,\beta,\, \varphi\in(0,2\pi)$. Then by taking $\beta=\pi,\varphi=0$ we obtain that the momenta are in the $(1,3)$ plane and the angle between momenta vectors is $\alpha+\gamma$ \cite{13}. The helicity bispinors summation given in \cite{13}, is included in the result :
\begin{eqnarray}\label{afcp}
\mathcal{P}_{e^{-}e^{+}}= \frac{\alpha^2Z^2k^{4/3}}{2(p^2+p'\,^2+2pp'\cos(\alpha+\gamma))^{2}\sinh^2(\pi k)}\frac{1}{p'\,^2}
\times\left\{
\begin{array}{cll}
\sin^2 \left(k\ln\left(\frac{p'}{p}\right)\right)\sin^2\left(\frac{\alpha+\gamma}{2}\right)&{\rm for}&\sigma=\sigma'\\
\cos^2 \left(k\ln\left(\frac{p'}{p}\right)\right)\cos^2\left(\frac{\alpha+\gamma}{2}\right)&{\rm for}&\sigma=-\sigma'.
\end{array}\right.
\end{eqnarray}
The trigonometric functions that contain the logarithm of the momenta ratio will be approximated as in the previous section, and we obtain in the case when $\sigma=\sigma'$ the equation that contain the momenta integrals:
\begin{eqnarray}\label{afcp}
\mathcal{P}_{\sigma=\sigma'}^{tot}= \frac{\alpha^2Z^2k^{10/3}}{2\sinh^2(\pi k)}\sin^2\left(\frac{\alpha+\gamma}{2}\right)\int_{P'_{min}/\omega}^{P'/\omega} dp'\int_{0}^{\infty} dp\frac{(p-p')^2}{(p^2+p'\,^2+2pp'\cos(\alpha+\gamma))^{2}}.
\end{eqnarray}
In the above equation we cutoff the integral after $p'$ since the integrals contain both infrared and ultraviolet divergences and this is the consequence of using the Coulomb potential which gives this kind of behaviour. The limits of the $p'$  integral will be between two values of the adimensional factors $\frac{P_{min}'}{\omega}$ and $\frac{P'}{\omega}$. The angle between momenta vectors are fixed and the final results given in Appendix, contain the total probability computed for fermions moving on given directions:
\begin{eqnarray}\label{as1}
\mathcal{P}_{\sigma=\sigma'}^{tot}= \frac{\alpha^2Z^2k^{10/3}}{2\sinh^2(\pi k)}\ln\left(\frac{P'}{P'_{min}}\right)\times\left\{
\begin{array}{cll}
0\,&{\rm for}&\alpha+\gamma=0\\
\left(\frac{2\pi\sqrt{3}+4\pi-6\sqrt{3}-12}{48} \right)\,&{\rm for}&\alpha+\gamma=\pi/6\\
\left(\frac{4\pi\sqrt{3}-18}{36} \right)\,&{\rm for}&\alpha+\gamma=\pi/3\\
\frac{8}{10}\left(\frac{8\pi\sqrt{3}-18}{27}\right)\,&{\rm for}&\alpha+\gamma=2\pi/3\\
\left(\frac{20\pi-10\pi\sqrt{3}+6\sqrt{3}-12}{3}\right)\,&{\rm for}&\alpha+\gamma=5\pi/6.
\end{array}\right.
\end{eqnarray}
In the case when polarizations are such that $\sigma=-\sigma'$, the helicity is conserved and we obtain for the probability equation using only the first term of the cosine function expansion:
\begin{eqnarray}\label{afcp}
\mathcal{P}_{\sigma=-\sigma'}^{tot}= \frac{\alpha^2Z^2k^{4/3}}{2\sinh^2(\pi k)}\cos^2\left(\frac{\alpha+\gamma}{2}\right)\int_{P'_{min}/\omega}^{P'/\omega} dp'\int_{0}^{\infty} dp\frac{p^2}{(p^2+p'\,^2+2pp'\cos(\alpha+\gamma))^{2}}.
\end{eqnarray}
The momenta integrals that help us to establish the final equations for total probabilities are given in appendix and the results for various angles are included:
\begin{eqnarray}\label{as2}
\mathcal{P}_{\sigma=-\sigma'}^{tot}= \frac{\alpha^2Z^2k^{4/3}}{2\sinh^2(\pi k)}\ln\left(\frac{P'}{P'_{min}}\right)\times\left\{
\begin{array}{cll}
\frac{1}{3}\,&{\rm for}&\alpha+\gamma=0\\
\left(\frac{2\pi-3\sqrt{3}}{3} \right)\,&{\rm for}&\alpha+\gamma=\pi/6\\
\frac{3}{4}\left(\frac{4\pi\sqrt{3}-9}{27} \right)\,&{\rm for}&\alpha+\gamma=\pi/3\\
\frac{1}{4}\left(\frac{8\pi\sqrt{3}+9}{27}\right)\,&{\rm for}&\alpha+\gamma=2\pi/3\\
\frac{1}{16}\left(\frac{10\pi\sqrt{3}+3\sqrt{3}}{3}\right)\,&{\rm for}&\alpha+\gamma=5\pi/6\\
0\,&{\rm for}&\alpha+\gamma=\pi.
\end{array}\right.
\end{eqnarray}

In the limit $k>>1$ the total probabilities equations (\ref{as1}) and (\ref{as2}) reduce to :
\begin{eqnarray}\label{l1}
\mathcal{P}_{\sigma=\sigma'}^{tot}|_{k>>1}\simeq \frac{\alpha^2Z^2\left(\frac{m}{\omega}\right)^{10/3}e^{-2\pi \frac{m}{\omega}}}{2}\ln\left(\frac{P'}{P'_{min}}\right)\times\left\{
\begin{array}{cll}
0\,&{\rm for}&\alpha+\gamma=0\\
\left(\frac{2\pi\sqrt{3}+4\pi-6\sqrt{3}-12}{48} \right)\,&{\rm for}&\alpha+\gamma=\pi/6\\
\left(\frac{4\pi\sqrt{3}-18}{36} \right)\,&{\rm for}&\alpha+\gamma=\pi/3\\
\frac{8}{10}\left(\frac{8\pi\sqrt{3}-18}{27}\right)\,&{\rm for}&\alpha+\gamma=2\pi/3\\
\left(\frac{20\pi-10\pi\sqrt{3}+6\sqrt{3}-12}{3}\right)\,&{\rm for}&\alpha+\gamma=5\pi/6,
\end{array}\right.
\end{eqnarray}

\begin{eqnarray}\label{l2}
\mathcal{P}_{\sigma=-\sigma'}^{tot}|_{k>>1}\simeq \frac{\alpha^2Z^2\left(\frac{m}{\omega}\right)^{4/3}e^{-2\pi\frac{m}{\omega}}}{2}\ln\left(\frac{P'}{P'_{min}}\right)\times\left\{
\begin{array}{cll}
\frac{1}{3}\,&{\rm for}&\alpha+\gamma=0\\
\left(\frac{2\pi-3\sqrt{3}}{3} \right)\,&{\rm for}&\alpha+\gamma=\pi/6\\
\frac{3}{4}\left(\frac{4\pi\sqrt{3}-9}{27} \right)\,&{\rm for}&\alpha+\gamma=\pi/3\\
\frac{1}{4}\left(\frac{8\pi\sqrt{3}+9}{27}\right)\,&{\rm for}&\alpha+\gamma=2\pi/3\\
\frac{1}{16}\left(\frac{10\pi\sqrt{3}+3\sqrt{3}}{3}\right)\,&{\rm for}&\alpha+\gamma=5\pi/6\\
0\,&{\rm for}&\alpha+\gamma=\pi.
\end{array}\right.
\end{eqnarray}
Equations (\ref{l1}) and (\ref{l2}) shows that the probabilities in the case $m>>\omega$ become small because of the factor $e^{-2\pi \frac{m}{\omega}}$. In the Minkowski limit $k\rightarrow\infty$ the total probabilities given above are zero and this corresponds to the well known result that spontaneous particle generation in Coulomb field in Minkowski theory is a forbidden by the energy and momentum conservation laws.

\subsection{Graphical analysis}
In this subsection the total probabilities given in equations (\ref{as1}) and (\ref{as2}) are studied graphically in terms of parameter $k$. In all graphs we set the values for the ratio $\frac{P'}{P'_{min}}>1$ and the natural logarithm will contribute with a numerical factor. Our graphical analysis prove that if we take big values for the ratio from logarithm $\frac{P'}{P'_{min}}\sim10^8$ the variation in probabilities is small if we compare with the probabilities obtained when $\frac{P'}{P'_{min}}\sim2,3,4...$ and this can be checked for all graphs that follows in our analysis. Thus we can restrict to take only small values for $\frac{P'}{P'_{min}}$.

\begin{figure}[h!t]
\includegraphics[scale=0.45]{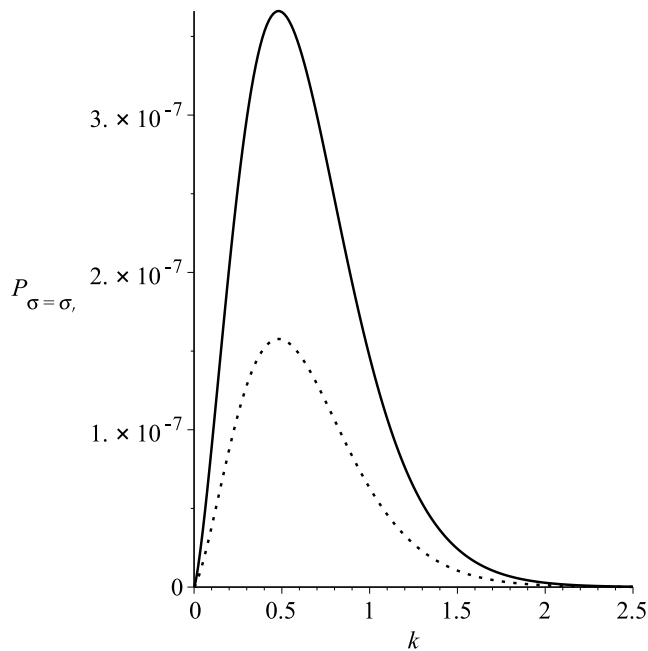}
\caption{Total probability as a function of parameter $k$ for $\alpha+\gamma=\frac{\pi}{6}$, in the case $\sigma=\sigma'$, solid line is for $\frac{P'}{P'_{min}}=5$ and point line is for $\frac{P'}{P'_{min}}=2$.}
\label{f11}
\end{figure}

\begin{figure}[h!t]
\includegraphics[scale=0.45]{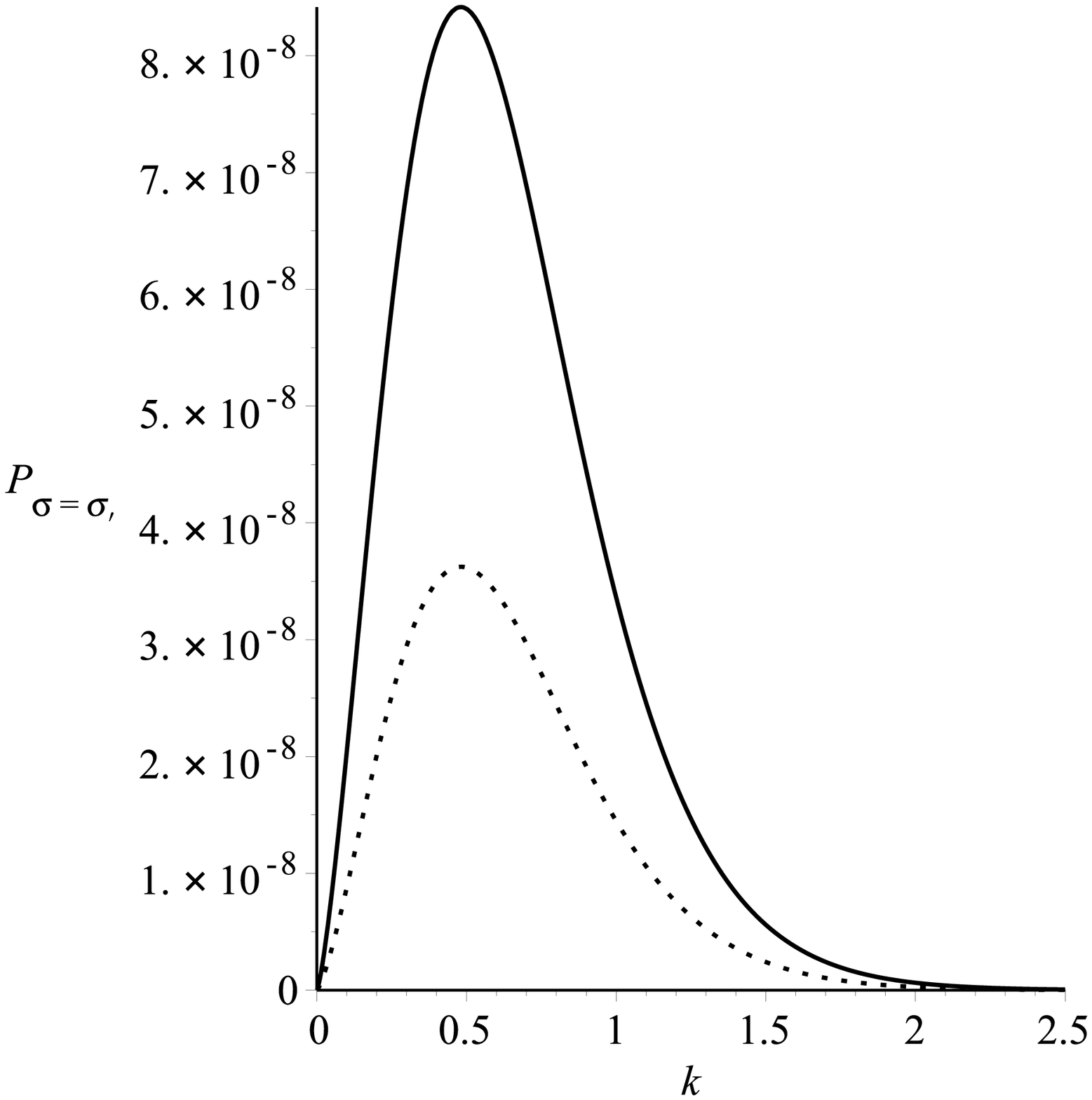}
\caption{Total probability as a function of parameter $k$ for $\alpha+\gamma=\frac{\pi}{3}$, in the case $\sigma=\sigma'$, solid line is for $\frac{P'}{P'_{min}}=5$ and point line is for $\frac{P'}{P'_{min}}=2$.}
\label{f12}
\end{figure}

\begin{figure}[h!t]
\includegraphics[scale=0.45]{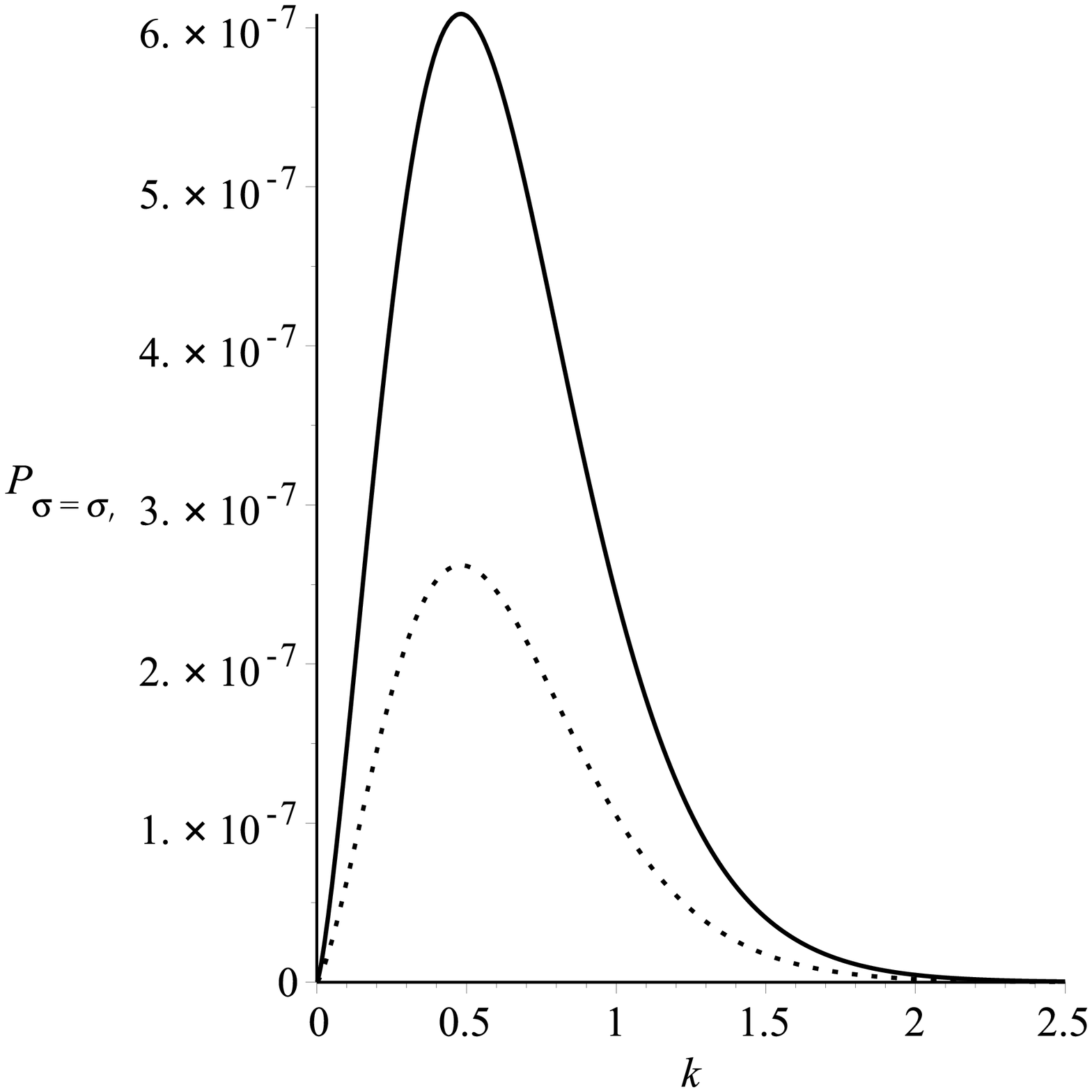}
\caption{Total probability as a function of parameter $k$ for $\alpha+\gamma=\frac{2\pi}{3}$, in the case $\sigma=\sigma'$, solid line is for $\frac{P'}{P'_{min}}=5$ and point line is for $\frac{P'}{P'_{min}}=2$.}
\label{f13}
\end{figure}

\begin{figure}[h!t]
\includegraphics[scale=0.45]{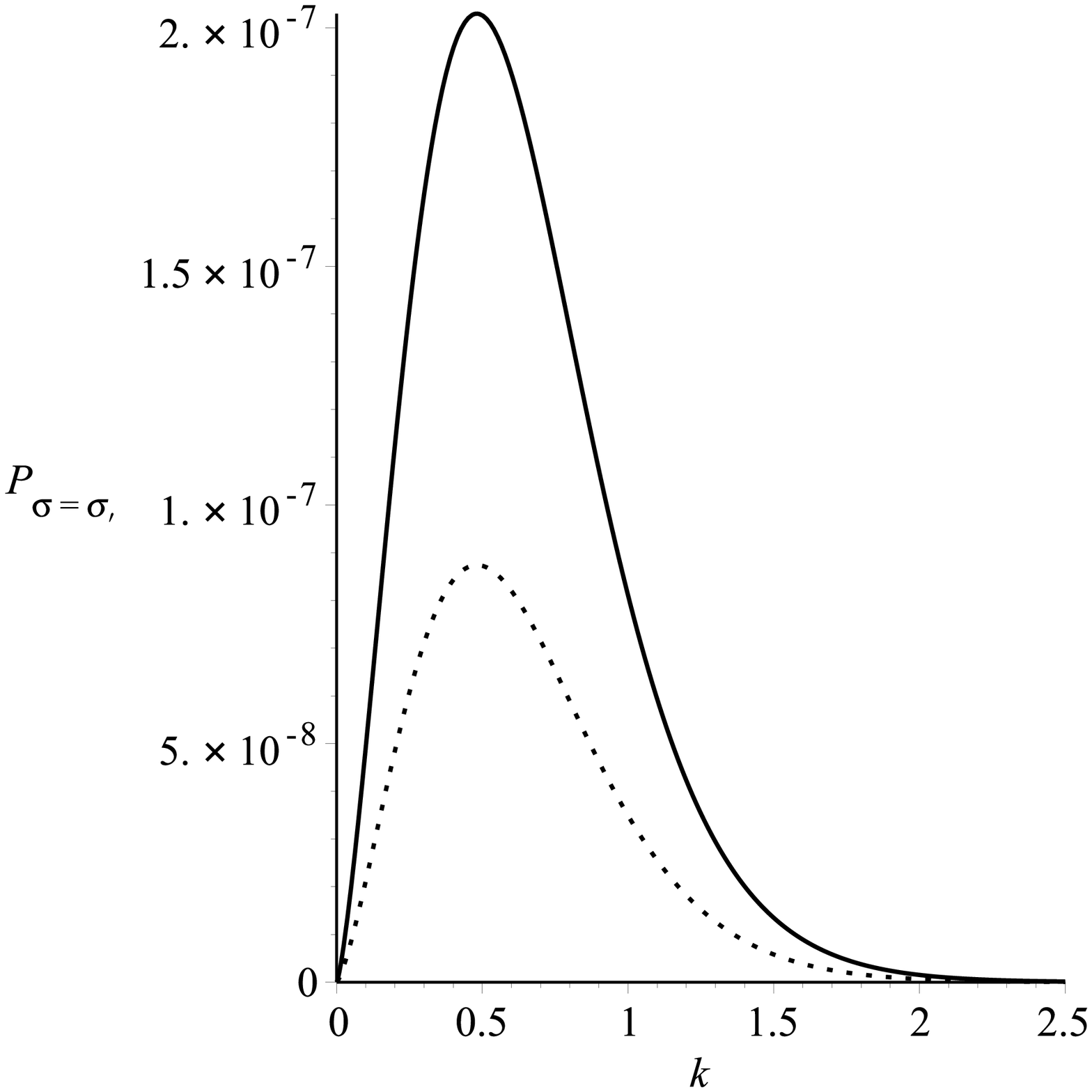}
\caption{Total probability as a function of parameter $k$ for $\alpha+\gamma=\frac{5\pi}{6}$, in the case $\sigma=\sigma'$, solid line is for $\frac{P'}{P'_{min}}=5$ and point line is for $\frac{P'}{P'_{min}}=2$.}
\label{f14}
\end{figure}

\begin{figure}[h!t]
\includegraphics[scale=0.45]{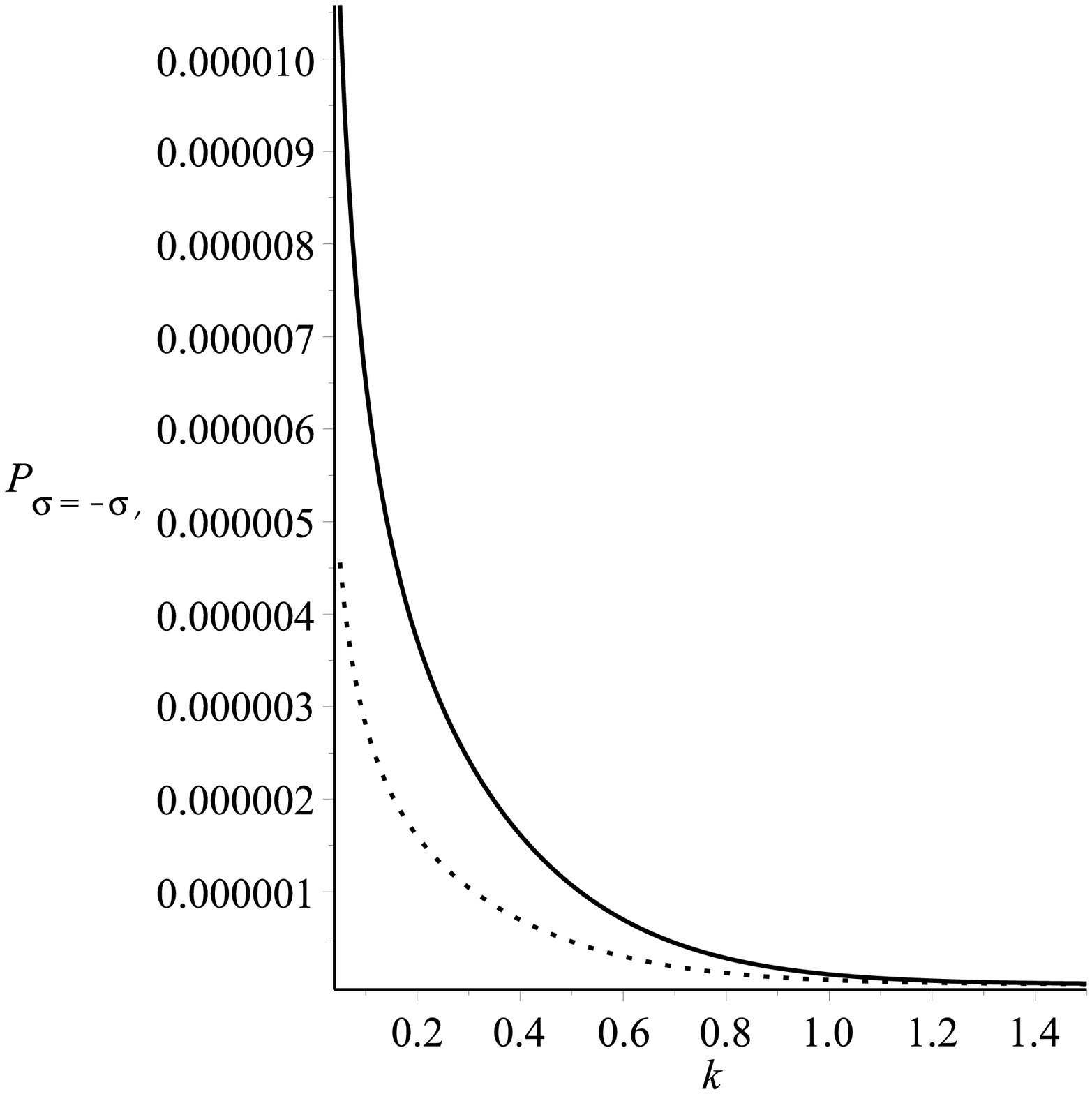}
\caption{Total probability as a function of parameter $k$ for $\alpha+\gamma=0$, in the case $\sigma=-\sigma'$, solid line is for $\frac{P'}{P'_{min}}=5$ and point line is for $\frac{P'}{P'_{min}}=2$.}
\label{f15}
\end{figure}

\begin{figure}[h!t]
\includegraphics[scale=0.45]{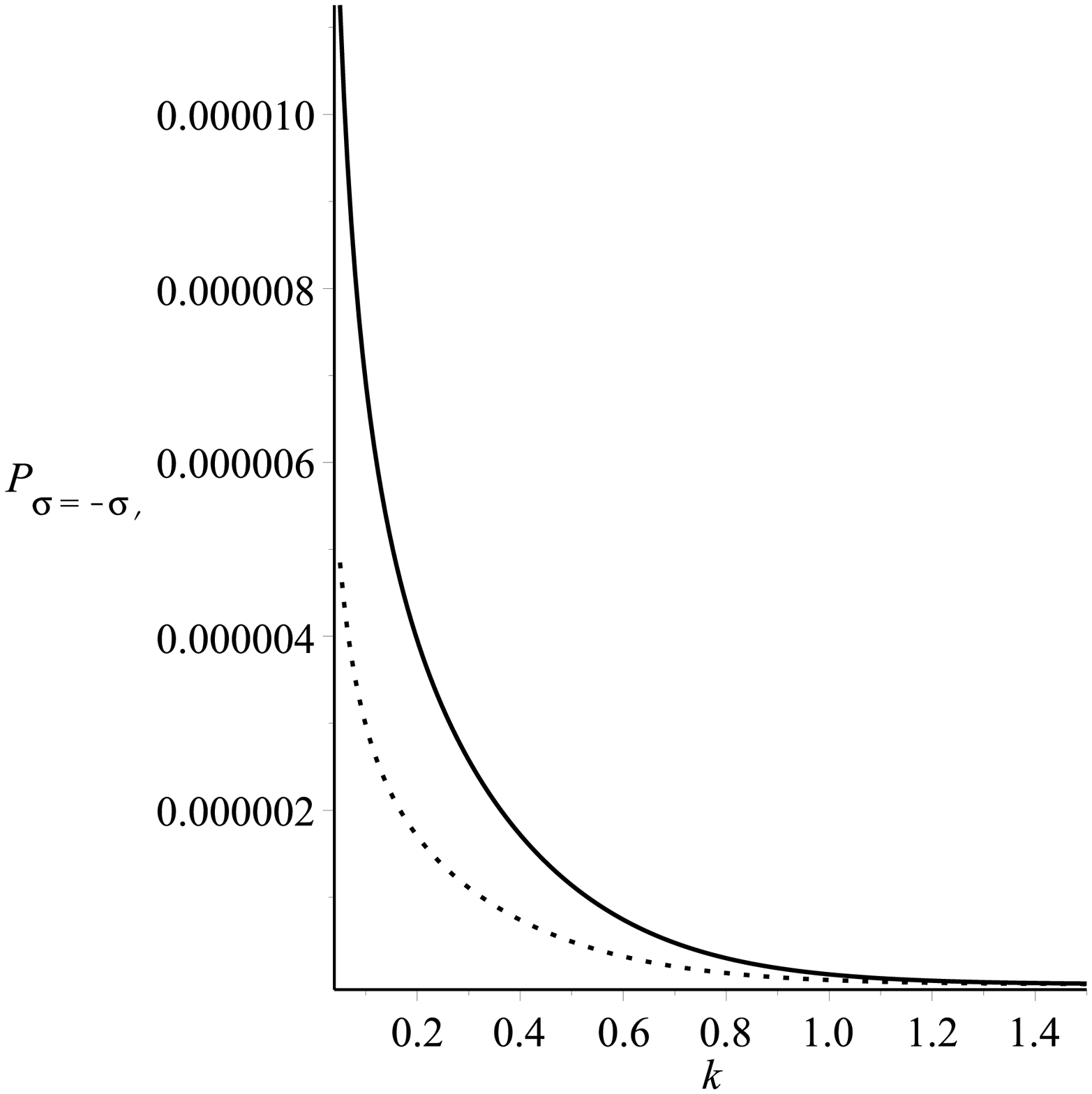}
\caption{Total probability as a function of parameter $k$ for $\alpha+\gamma=\frac{\pi}{3}$, in the case $\sigma=-\sigma'$, solid line is for $\frac{P'}{P'_{min}}=5$ and point line is for $\frac{P'}{P'_{min}}=2$.}
\label{f16}
\end{figure}

\begin{figure}[h!t]
\includegraphics[scale=0.45]{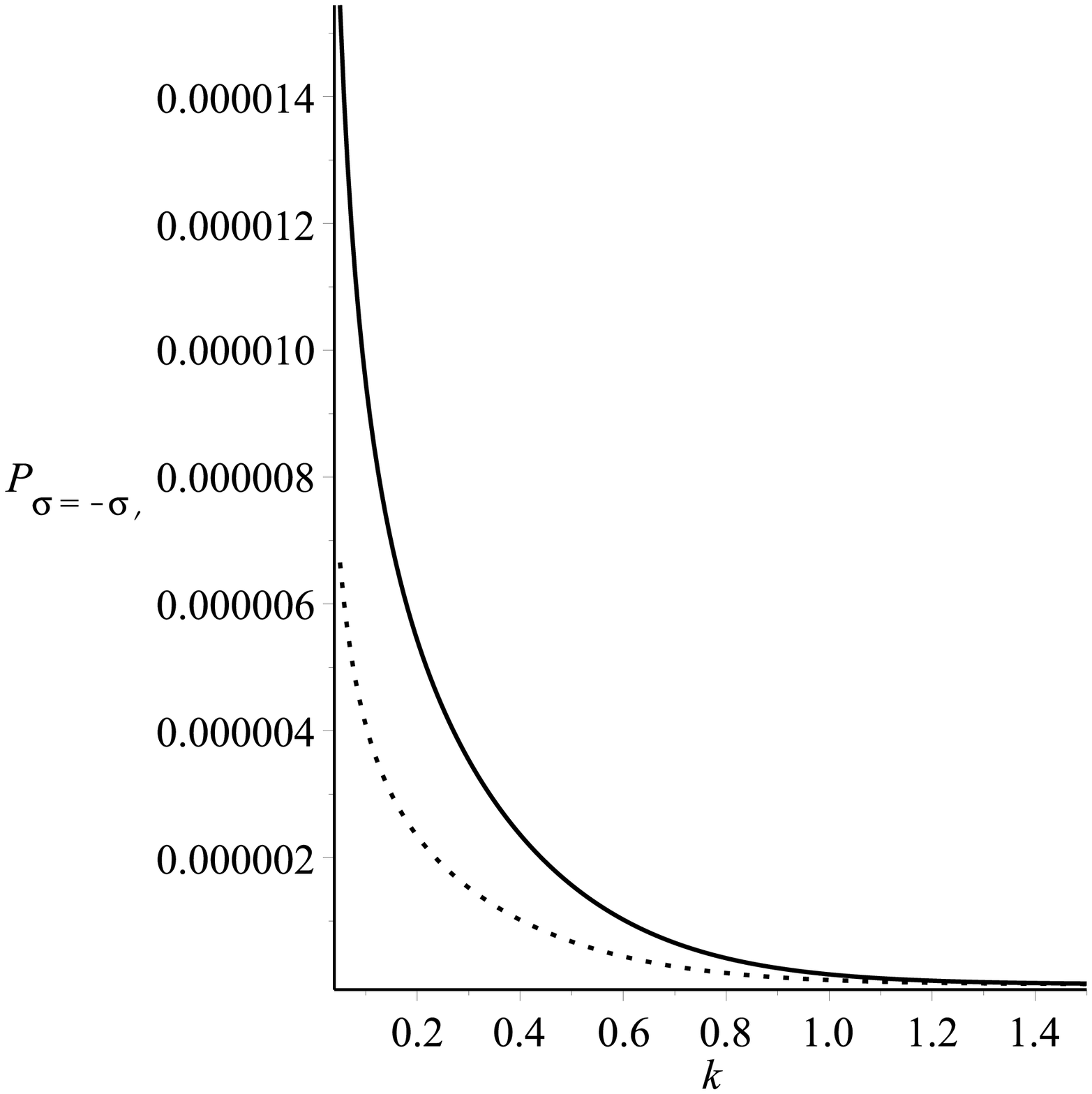}
\caption{Total probability as a function of parameter $k$ for $\alpha+\gamma=\frac{2\pi}{3}$, in the case $\sigma=-\sigma'$, solid line is for $\frac{P'}{P'_{min}}=5$ and point line is for $\frac{P'}{P'_{min}}=2$.}
\label{f17}
\end{figure}

\begin{figure}[h!t]
\includegraphics[scale=0.45]{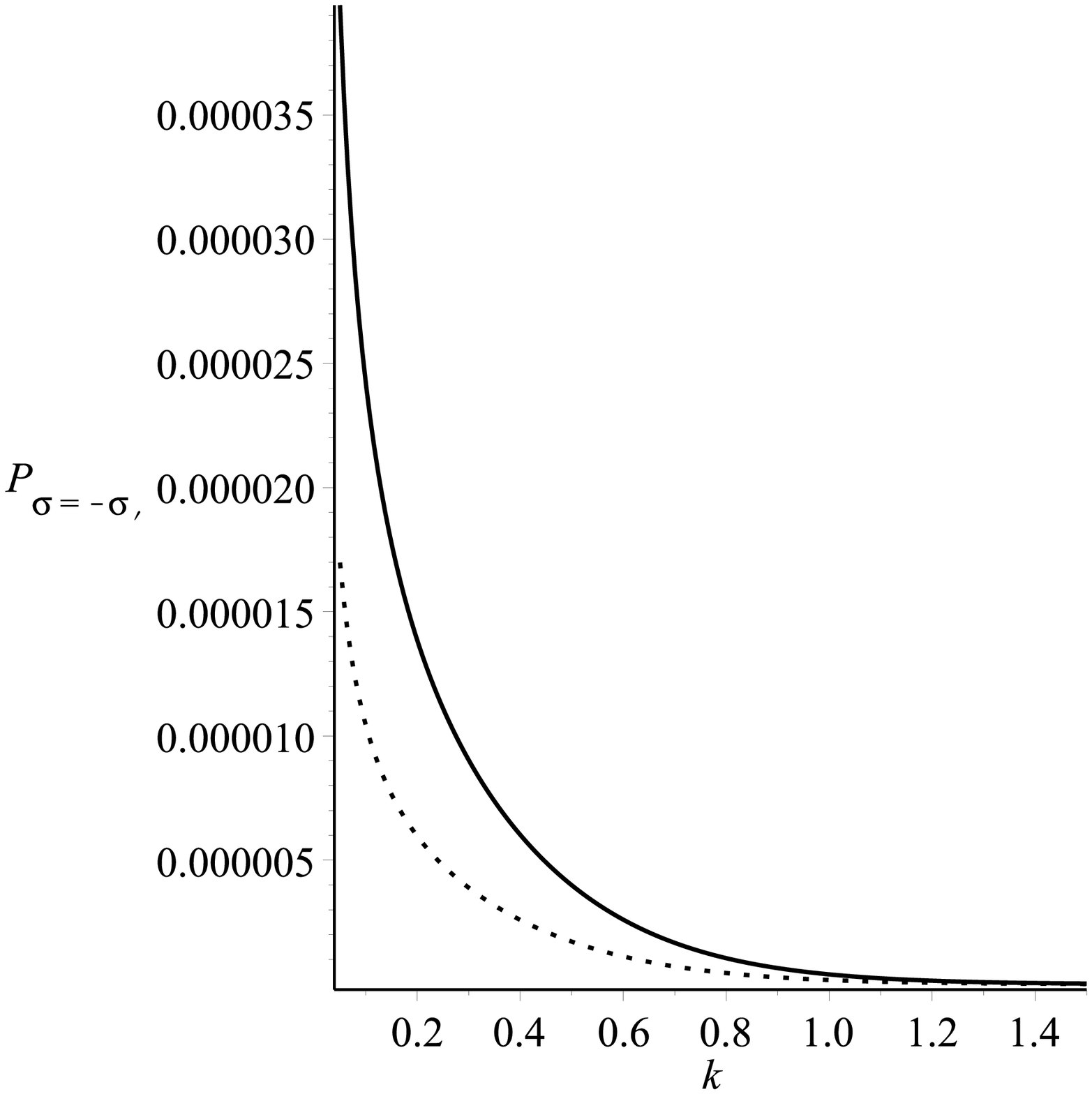}
\caption{Total probability as a function of parameter $k$ for $\alpha+\gamma=\frac{5\pi}{6}$, in the case $\sigma=-\sigma'$, solid line is for $\frac{P'}{P'_{min}}=5$ and point line is for $\frac{P'}{P'_{min}}=2$.}
\label{f18}
\end{figure}

In the case when helicity is not conserved, Figs.(\ref{f11})-(\ref{f14}) we observe that the total probabilities are zero for $k=0$ and preserve the behaviour of the probability density obtained in \cite{13}. The graphs Figs.(\ref{f15})-(\ref{f18}) contain the total probability dependence on parameter $k$ for various angles in the case when helicity is conserved. In the Minkowski limit our graphs prove that the probabilities are zero and this result confirms the results obtained analytically in the previous section.

By using the approximation given in equation (\ref{apx}), we obtain that the total probabilities dependence on parameter $k$ preserve the general behaviour of the probability density as obtained in \cite{13}. We mention that the results given in graphs (\ref{f1})-(\ref{f18}) were obtained using Maple.

The results obtained in  Figs.(\ref{f11})-(\ref{f18}) also show that the helicity conserving processes are favoured since the total probabilities are sensibly larger in this case (see Figs.(\ref{f15})-(\ref{f18})). However for a complete understanding of the results one needs to translate all the results obtained in this section in measurable quantities such as the number of fermions.

\newpage
\section{Number of fermions}
The topic of this section is related to the physical significance of the total probabilities computed above. A closer look to the final equations for total probabilities prove that these are dimensionless or can be made dimensionless in the case of fermions production in magnetic field. This means that our results can be translated in the number of produced particles. Here an important observation is related to the fact that we do not have a delta Dirac function dependent of momenta that conserve the momentum vector, instead we have a Heaviside step function that tell us that the momenta modulus of the produced particles are not equal, that means that the momentum conservation law is broken. In the case of fermion pair production in Coulomb field the number of fermions is contained in equations (\ref{as1}), (\ref{as2}), and these quantities were computed by summing after the polarizations $\sigma,\sigma'$ and integrating after the momenta $p,p\,'$ . For obtaining the number of particles per unit of proper volume $V$, we will divide the total probability to unit of proper volume, following the general results obtained in \cite{18}, where the number of particles was computed by using perturbative approach:
\begin{equation}
n=\frac{\mathcal{P}^{tot}}{V}.
\end{equation}
Then in the case of pair production in Coulomb field the total number of fermions is obtained by summing the contributions from the processes that conserve the helicity and processes that broken the helicity conservation law:
\begin{equation}
n_{Coul}=\frac{\mathcal{P}_{\sigma=\sigma'}^{tot}}{V}+\frac{\mathcal{P}_{\sigma=-\sigma'}^{tot}}{V}.
\end{equation}
For example the result for the total number of produced fermions in unit of proper volume when the angle between momenta vectors is fixed at
$\alpha+\gamma=\frac{\pi}{3}$ give:
\begin{equation}
n_{Coul}|_{\frac{\pi}{3}}=\frac{\alpha^2Z^2k^{4/3}}{2V\sinh^2(\pi k)}\ln\left(\frac{P'}{P'_{min}}\right)\left[k^2\left(\frac{4\pi\sqrt{3}-18}{36}\right)+\frac{3}{4}\left(\frac{4\pi\sqrt{3}-9}{27}\right)\right].
\end{equation}
while for $\alpha+\gamma=\frac{5\pi}{6}$ we obtain:
\begin{equation}
n_{Coul}|_{\frac{5\pi}{6}}=\frac{\alpha^2Z^2k^{4/3}}{2V\sinh^2(\pi k)}\ln\left(\frac{P'}{P'_{min}}\right)\left[k^2\left(\frac{20\pi-10\pi\sqrt{3}+6\sqrt{3}-12}{3}\right)+\frac{1}{16}\left(\frac{10\pi\sqrt{3}+3\sqrt{3}}{3}\right)\right].
\end{equation}
All the expressions for total number of produced fermions can be established by using equations (\ref{as1}) and (\ref{as2}). Our graphs Figs. (\ref{f11})-(\ref{f18}) contain the number of particles (on ordinate) in unit of proper volume. For comparing the number of fermions for different angles we give also the number equation for $\alpha+\gamma=0$, which give contribution only for $\sigma=-\sigma'$:
\begin{equation}
n_{Coul}|_{0}=\frac{\alpha^2Z^2k^{4/3}}{2V\sinh^2(\pi k)}\ln\left(\frac{P'}{P'_{min}}\right)\frac{1}{3}.
\end{equation}
\begin{figure}[h!t]
\includegraphics[scale=0.45]{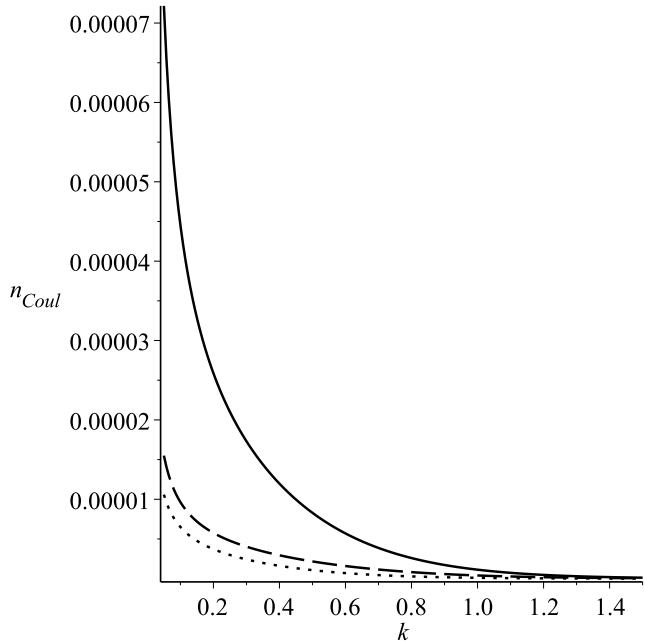}
\caption{The number of fermions in Coulomb field in terms of parameter $k$ for $\alpha+\gamma=\frac{5\pi}{6}$ solid line, $\alpha+\gamma=\frac{2\pi}{3}$ dashed line and $\alpha+\gamma=0$ the point line.}
\label{f19}
\end{figure}
The result presented in Fig. (\ref{f19}) proves that number of fermions increases with the angle between the momenta vectors ($\alpha+\gamma$), and is minimum at $\alpha+\gamma=0$, this being the situation when the fermions move along the same directions and the momenta are parallel, having as result the pair annihilation. The number of fermions increase as the angle between momenta vectors approaches $\pi$.

Let us discuss now the problem related to the number of produced fermions in dipole magnetic field. Here the first observation is related to the fact that the amplitude and probability contain the dipole magnetic moment $\vec{\mathcal{M}}$ and for obtaining a number we have to divide the probability by a physical quantity that has the same units. As in previous case for obtaining the number of fermions we need to divide the total probability to the volume and sum the contributions from the helicity conserving processes and helicity non-conserving processes. The total number of produced fermions in unit of proper volume in dipole magnetic field when the angle between momenta vectors is fixed at $\beta-\varphi=\frac{2\pi}{3}$ will be proportional with:
\begin{eqnarray}
n_{mag}&\sim&\frac{\alpha\mathcal{M}^2}{2\pi^{3}V\cosh^2(\pi k)}\left(\frac{P'}{\omega}\right)^2\frac{3}{4}
\biggl\{k^2\left(\frac{8\pi\sqrt{3}}{9}+ \frac{18}{27}\right)+\frac{1}{3}\nonumber\\
&& + \frac{4}{9}\left(\ln\left(\frac{P'}{\omega}\right)-\frac{1}{2}\right)- \frac{2}{9}\left(\ln\left(\frac{P'}{\omega}\right)^2-1\right)\biggl\},
\end{eqnarray}
while in the case $\beta-\varphi=\frac{\pi}{6}$ we obtain:
\begin{eqnarray}
n_{mag}&\sim&\frac{\alpha\mathcal{M}^2}{2\pi^{3}V\cosh^2(\pi k)}\left(\frac{P'}{\omega}\right)^2\biggl\{k^2\left(\frac{1}{2}+\frac{\sqrt{3}}{4}\right)
\left(\frac{4\pi}{3}-\frac{2\pi\sqrt{3}}{3} -\frac{6\sqrt{3}}{3}+4\right)\nonumber\\
&& + \left(\frac{1}{2}-\frac{\sqrt{3}}{4}\right)\left(\frac{2\pi+2\pi\sqrt{3}}{-7+4\sqrt{3}}+ \frac{20\pi+27\sqrt{3}-48-12\pi\sqrt{3}}{168\sqrt{3}-291}\right) \nonumber\\
&&+\left(\frac{1}{2}-\frac{\sqrt{3}}{4}\right)\left(\frac{48\sqrt{3}-84}{168\sqrt{3}-291}\right)\left(\ln\left(\frac{P'}{\omega}\right)-\frac{1}{2}\right)\nonumber\\
&&+\left(\frac{1}{2}-\frac{\sqrt{3}}{4}\right)\left(\frac{42-24\sqrt{3}}{168\sqrt{3}-291}\right)\left(\ln\left(\frac{P'}{\omega}\right)^2-1\right)\biggl\}.
\end{eqnarray}
For $\beta-\varphi=0$ the result is:
\begin{eqnarray}
n_{mag}&\sim&\frac{\alpha\mathcal{M}^2}{2\pi^{3}V\cosh^2(\pi k)}\left(\frac{P'}{\omega}\right)^2.
\end{eqnarray}
 Plotting the above equations we obtain the number dependence on parameter $k$:
\begin{figure}[h!t]
\includegraphics[scale=0.45]{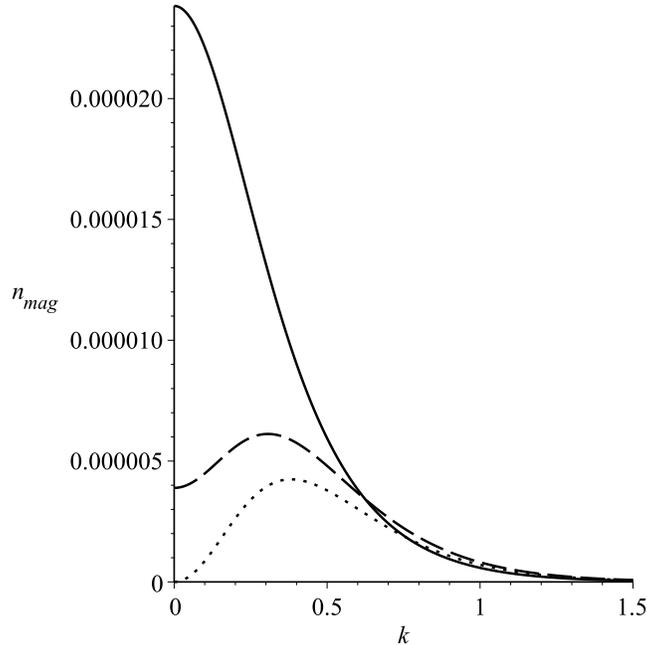}
\caption{The number of fermions in dipole magnetic field in terms of parameter $k$ for $\beta-\varphi=\frac{2\pi}{3}$ solid line, $\beta-\varphi=\frac{\pi}{3}$ dashed line and $\beta-\varphi=0$ the point line.}
\label{f20}
\end{figure}

The result obtained in Fig. (\ref{f20}) shows that the number of fermions increase as the angle between the momenta of the fermion and momenta of the anti-fermion increase and this confirm the above results obtained in the case of pair generation in Coulomb field.  For large $k$ the number of produced fermions have a exponential decrease in both cases studied in the present paper. In the Minkowski limit the number of fermions is zero as we expected since the perturbative processes of spontaneous particle generation in external fields are forbidden in flat geometry.

The results obtained here prove that the phenomenon of particle production by using perturbative methods can have as output the same quantities as in the case of cosmological particle production were the number densities are obtained \cite{bs,31,32,33}. The difference is that using the perturbative methods one can study all the interesting cases such as the case of particle generation at large expansion. The problem of particle generation at fields interactions received little attention and here we mention the important result obtained in \cite{18}. In \cite{18} the authors discuss the conditions in which the particle generation at fields interactions could become more important as the cosmological particle production and a few interesting examples were analysed and we mention that only massless fields were considered. Our results present for the first time how the number of massive fermions could be obtained from a perturbative calculation.

\section{Discussion}

The results obtained in this paper are related to the computations of the first order transition probabilities corresponding to the processes of fermion production in external fields in a de Sitter geometry. The methods used are based on perturbations that help us to compute S matrix elements corresponding to the de Sitter QED and this formalism was studied in \cite{7}. The advantage of using this method is related to the fact that the gravity dependence is preserved in our equations for probabilities and this allows us to consider the situation when the gravitational fields are strong or in other words to study the probabilities for the regime of large expansion from early universe.

We establish here the first result for the total probability of fermion pair production in dipole magnetic field and Coulomb field on de Sitter space-time. This method can be used further to compute the total probabilities for fermion generation in electric fields on de Sitter geometry. Remarkably is that the total probabilities preserve the general profile of the dependence in terms of parameter $k$, as the probability densities obtained in \cite{10,13}. Another important result is that the phenomenon of particle production is relevant only when the expansion parameter is large comparatively with the mass of the fermion. The results from flat space QED are recovered in the sense that the probabilities are vanishing in this limit. The number of fermions was computed, and we obtain that this quantity is dependent on the momenta orientation of the generated fermions.

\section{Appendix}

For  establishing the equation for transition amplitude and probability in terms of trigonometric functions, we use \cite{16,22}:
\begin{equation}\label{1}
a^{ix}=e^{ix\ln(a)}, a\in\mathcal{R}.
\end{equation}
The momentum integrals that help us to establish the final expression for the total probability of fermion production in magnetic field in the case $\sigma=\sigma'$ are:
\begin{eqnarray}
&&\int_0^{\infty}dp\frac{1}{(p+p\,')^2}=\frac{1}{p\,'},\nonumber\\
&&\int_0^{\infty}dp\frac{(p+p\,')^2}{(p^2+p\,'^2+\sqrt{3}pp\,')^2}=\frac{1}{p\,'}\left(\frac{4\pi}{3}-\frac{2\pi\sqrt{3}}{3} -\frac{6\sqrt{3}}{3}+4\right),\nonumber\\
&&\int_0^{\infty}dp\frac{(p+p\,')^2}{(p^2+p\,'^2+pp\,')^2}=\frac{1}{p\,'}\left(\frac{4\pi\sqrt{3}}{27}+ \frac{18}{27}\right),\nonumber\\
&&\int_0^{\infty}dp\frac{(p+p\,')^2}{(p^2+p\,'^2-pp\,')^2}=\frac{1}{p\,'}\left(\frac{8\pi\sqrt{3}}{27}+ \frac{18}{27}\right),\nonumber\\
&&\int_0^{\infty}dp\frac{(p+p\,')^2}{(p^2+p\,'^2-\sqrt{3}pp\,')^2}=\frac{1}{p\,'}\left(\frac{20\pi}{3}+\frac{10\pi\sqrt{3}}{3} +\frac{6\sqrt{3}}{3}+4\right).
\end{eqnarray}

For $\sigma=-\sigma'$ the momentum integrals for computing the total probability of fermion production in magnetic field are:
\begin{eqnarray}
&&\int_{0}^{\infty}dp\frac{p^2(p-p\,')^2}{\left(p+p'\right)^2(p^2+p\,'^2+\sqrt{3}pp\,')^2}=\frac{1}{p\,'}\left(\frac{2\pi+2\pi\sqrt{3}}{-7+4\sqrt{3}}+ \frac{20\pi+27\sqrt{3}-48-12\pi\sqrt{3}}{168\sqrt{3}-291}\right)\nonumber\\
&&+\frac{1}{p\,'}\left(\frac{48\sqrt{3}-84}{168\sqrt{3}-291}\ln(p\,')+\frac{42-24\sqrt{3}}{168\sqrt{3}-291}\ln(p\,'^2)\right),\nonumber\\
&&\int_{0}^{\infty}dp\frac{p^2(p-p\,')^2}{\left(p+p'\right)^2(p^2+p\,'^2+pp\,')^2}=\frac{1}{p\,'}\left( -\frac{8\pi\sqrt{3}}{9}+ 5+4\ln(p\,')-2\ln(p\,'^2)\right),\nonumber\\
&&\int_{0}^{\infty}dp\frac{p^2(p-p\,')^2}{\left(p+p'\right)^2(p^2+p\,'^2-pp\,')^2}=\frac{1}{p\,'}\left(\frac{1}{3}+\frac{4}{9}\ln(p\,')-\frac{2}{9}\ln(p\,'^2)\right).
\end{eqnarray}
\begin{eqnarray}
&&\int_{0}^{P'/\omega}dp\,'p\,'\ln(p\,')=\frac{1}{2}\left(\frac{P'}{\omega}\right)^2\left(\ln\left(\frac{P'}{\omega}\right)-\frac{1}{2}\right),\nonumber\\
&&\int_{0}^{P'/\omega}dp\,'p\,'\ln(p\,'^2)=\frac{1}{2}\left(\frac{P'}{\omega}\right)^2\left(\ln\left(\frac{P'}{\omega}\right)^2-1\right),\nonumber\\
&&\int_{P'_{min}/\omega}^{P'/\omega}dp\,'\frac{1}{p\,'}=\ln\left(\frac{P'}{P'_{min}}\right).
\end{eqnarray}
In the case of fermion production in Coulomb field the momenta integrals used for obtaining the total probabilities are:
\begin{eqnarray}
&&\int_{0}^{\infty}dp\frac{(p-p\,')^2}{(p^2+p\,'^2+\sqrt{3}pp\,')^2}=\frac{1}{p\,'}\left(\frac{4\pi+2\pi\sqrt{3}-6\sqrt{3}-12}{3}\right),\nonumber\\
&&\int_{0}^{\infty}dp\frac{(p-p\,')^2}{(p^2+p\,'^2+pp\,')^2}=\frac{1}{p\,'}\left( \frac{4\pi\sqrt{3}-18}{9}\right),\nonumber\\
&&\int_{0}^{\infty}dp\frac{p^2}{(p^2+p\,'^2+2pp\,')^2}=\frac{1}{3p\,'},\nonumber\\
&&\int_{0}^{\infty}dp\frac{p^2}{(p^2+p\,'^2+\sqrt{3}pp\,')^2}=\frac{1}{3p\,'}(2\pi-3\sqrt{3}),\nonumber\\
&&\int_{0}^{\infty}dp\frac{p^2}{(p^2+p\,'^2+pp\,')^2}=\frac{1}{27p\,'}(4\pi\sqrt{3}-9).
\end{eqnarray}

\end{document}